\documentclass[10pt,journal,compsoc]{IEEEtran}

\usepackage[utf8]{inputenc}
\usepackage[british]{babel}
\usepackage{csquotes}
\usepackage{graphicx}
\usepackage{subfig}
\usepackage[table]{xcolor}
\usepackage{xcolor}
\usepackage{url}
\usepackage{booktabs}
\usepackage{float}
\usepackage{amsmath}
\usepackage{amssymb}
\usepackage[export]{adjustbox}
\usepackage{listings}
\usepackage{color}
\usepackage{tcolorbox}
\usepackage{multirow}
\usepackage[T1]{fontenc}
\usepackage{textcomp}
\usepackage{rotating} 
\usepackage{adjustbox} 
\usepackage{lineno, blindtext}
\usepackage{makecell}
\usepackage{tabularx}

\usepackage[normalem]{ulem} 
\usepackage{xspace} 
\usepackage{xcolor}
\definecolor{colgray}{gray}{0.75}

\usepackage{ifthen}
\usepackage{amssymb}
\newboolean{showcomments}
\setboolean{showcomments}{true} 
\ifthenelse{\boolean{showcomments}}
  {\newcommand{\nb}[2]{
    \fcolorbox{gray}{yellow}{\bfseries\sffamily\scriptsize#1}
    {$\blacktriangleright$#2$\blacktriangleleft$}
   }
   
  }
  {\newcommand{\nb}[2]{}
   
  }

\newcommand{\dependencyREQUIRES}{Requires\xspace}
\newcommand{\dependencyOR}{Or\xspace}
\newcommand{\dependencyXOR}{OnlyOne\xspace}
\newcommand{\dependencyXNOR}{AllOrNone\xspace}
\newcommand{\dependencyNAND}{ZeroOrOne\xspace}
\newcommand{\dependencyARITHMETIC}{Arithmetic/Relational\xspace}
\newcommand{\dependencyCOMPLEX}{Complex\xspace}

\newcommand{\idl}{IDL\xspace}

\newcommand{\noperations}{nine\xspace}

\newcommand{\ntestcases}{178\xspace}

\newcommand{\consistency}{Consistent specification\xspace}
\newcommand{\deadParameter}{Dead parameter\xspace}
\newcommand{\falseOptional}{False optional\xspace}
\newcommand{\validSpecification}{Valid specification\xspace}
\newcommand{\validRequest}{Valid request\xspace}
\newcommand{\validPartialRequest}{Valid partial request\xspace}
\newcommand{\randomRequest}{Random request\xspace}

\newcommand{\allRequests}{All requests\xspace}
\newcommand{\numberRequests}{Number of requests\xspace}

\definecolor{lightgray}{rgb}{.9,.9,.9}
\definecolor{darkgray}{rgb}{.4,.4,.4}
\definecolor{purple}{rgb}{0.65, 0.12, 0.82}
\definecolor{darkpurple}{rgb}{0.39, 0.10, 0.40}
\definecolor{backcolour}{rgb}{0.95,0.95,0.92}
\definecolor{codegray}{rgb}{0.5,0.5,0.5}
\definecolor{codegreen}{rgb}{0,0.6,0}

\lstdefinelanguage{IDL}{
  keywords={IF, THEN, AND, OR, NOT, Or, ZeroOrOne, OnlyOne, AllOrNone, LIKE},
  sensitive=true,
  keywordstyle=\color{darkpurple}\bfseries,
  ndkeywordstyle=\color{codegreen},
  identifierstyle=\color{black},
  comment=[l]{//},
  morecomment=[s]{/*}{*/},
  commentstyle=\color{codegreen}\ttfamily,
  stringstyle=\color{red}\ttfamily,
  morestring=[b]',
  morestring=[b]",
}

\lstdefinelanguage{Grammar}{
  keywords={Model:, Dependency:, RelationalDependency:, Param:, ArithmeticDependency:, Operation:, OperationContinuation:, ConditionalDependency:, Predicate:, Clause:, Term:, ParamValueRelation:, ClauseContinuation:, PredefinedDependency:, RelationalOperator:, ArithmeticOperator:, Not:, LogicalOperator:},
  sensitive=true,
  keywordstyle=\color{blue}\bfseries,
  ndkeywords={ID:, STRING:, DOUBLE:, PATTERN_STRING:, BOOLEAN:},
  ndkeywordstyle=\color{purple},
  identifierstyle=\color{black},
  comment=[l]{//},
  morecomment=[s]{/*}{*/},
  commentstyle=\color{codegreen}\ttfamily,
  stringstyle=\color{red}\ttfamily,
  morestring=[b]',
  morestring=[b]",
  alsodigit={:}
}

\lstdefinelanguage{MAP}{
  keywords={map, \{, \}, V, D, C},
  sensitive=true,
  keywordstyle=\color{black}\bfseries,
  ndkeywordstyle=\color{codegreen},
  identifierstyle=\color{black},
  comment=[l]{//},
  morecomment=[s]{/*}{*/},
  commentstyle=\color{codegreen}\ttfamily,
  stringstyle=\color{red}\ttfamily,
  morestring=[b]',
  morestring=[b]",
}

\lstdefinelanguage{YAML}{
  keywords={paths:, search:, get:, operationId:, parameters:, name:, x-dependencies:},
  sensitive=true,
  keywordstyle=\color{black},
  identifierstyle=\color{blue},
  comment=[l]{\#},
  commentstyle=\color{codegreen}\ttfamily,
  stringstyle=\color{red}\ttfamily,
  morestring=[b]',
  morestring=[b]",
  alsodigit={:, -}
}

\lstdefinestyle{listingtop}{
  float=tp,
  floatplacement=tbp,
  belowcaptionskip=-0.5cm,
}

\ifCLASSOPTIONcompsoc
  \usepackage[nocompress]{cite}
\else
  \usepackage{cite}
\fi
\ifCLASSINFOpdf
\else
\fi
\usepackage{url}

\begin{document}

\title{Specification and Automated Analysis of Inter-Parameter Dependencies in Web APIs}

%
%

\author{Alberto Martin-Lopez,
        Sergio Segura,
        Carlos Müller,
        and Antonio Ruiz-Cortés
\IEEEcompsocitemizethanks{\IEEEcompsocthanksitem A. Martin-Lopez, S. Segura, C. Müller, and A. Ruiz-Cortés are with the Department of Computer Languages and Systems, Universidad de Sevilla, Seville,
41012, Spain.\protect\\
E-mail: \{amarlop, sergiosegura, cmuller, aruiz\}@us.es}
}
\IEEEtitleabstractindextext{%
\begin{abstract}
Web services often impose inter-parameter dependencies that restrict the way in which two or more input parameters can be combined to form valid calls to the service. Unfortunately, current specification languages for web services like the OpenAPI Specification (OAS) provide no support for the formal description of such dependencies, which makes it hardly possible to automatically discover and interact with services without human intervention. In this article, we present an approach for the specification and automated analysis of inter-parameter dependencies in web APIs. We first present a domain-specific language, called \emph{Inter-parameter Dependency Language} (IDL), for the specification of dependencies among input parameters in web services. Then, we propose a mapping to translate an \idl document into a constraint satisfaction problem (CSP), enabling the automated analysis of \idl specifications using standard CSP-based reasoning operations. Specifically, we present a catalogue of \noperations analysis operations on \idl documents allowing to compute, for example, whether a given request satisfies all the dependencies of the service. Finally, we present a tool suite including an editor, a parser, an OAS extension, a constraint programming-aided library, and a test suite supporting \idl specifications and their analyses. Together, these contributions pave the way for a new range of specification-driven applications in areas such as code generation and testing.

\end{abstract}

}

\maketitle

\IEEEdisplaynontitleabstractindextext

%
\IEEEpeerreviewmaketitle

\section{Introduction}
\label{sec-intro}

Web Application Programming Interfaces (APIs) allow systems to interact with each other over the network, typically using web services~\cite{jacobson11-book,richardson13-book}. Web APIs are rapidly proliferating as the cornerstone for software integration enabling new consumption models such as mobile, social, Internet of Things (IoT), or cloud applications. 
Many companies are also exposing their existing assets as private APIs, fostering reusability, integration, and innovation within the boundaries of their own companies~\cite{jacobson11-book,jacobson14-oscon}. 
Popular API directories such as ProgrammableWeb~\cite{programmableweb} and RapidAPI~\cite{rapidapi} currently index over 22K and 10K web APIs, respectively, from multiple domains such as shopping, finances, social networks, or telephony.


Modern web APIs typically adhere to the REpresentational State Transfer (REST) architectural style, being referred to as RESTful web APIs~\cite{fielding00-phd}. \emph{RESTful web APIs} are decomposed into multiple web services, where each service implements one or more create, read, update, or delete (CRUD) operations over a resource (e.g., an invoice in the PayPal API), typically through HTTP interactions. RESTful APIs are commonly described using languages such as the OpenAPI Specification (OAS)~\cite{oai}, originally created as a part of the Swagger tool suite~\cite{swagger}, or the RESTful API Modeling Language (RAML)~\cite{raml}. These languages are designed to provide a structured description of a RESTful web API that allows both humans and computers to discover and understand the capabilities of a service without requiring access to the source code or additional documentation. Once an API is described in an OAS document, for example, the specification can be used to generate documentation, code (clients and servers), or even basic automated test cases~\cite{swagger}. In this article, we focus on RESTful web APIs and OAS as the arguable standards for web integration.
In what follows, we will use the terms RESTful web API, web API, or simply API interchangeably.

Web services often impose dependency constraints that restrict the way in which two or more input parameters can be combined to form valid calls to the service, we call these \emph{inter-parameter dependencies} (or simply \emph{dependencies} henceforth). For instance, it is common that the inclusion of a parameter requires or excludes---and therefore depends on---the use of some other parameter or group of parameters. As an example, the documentation of the Twilio API \cite{twilio-api} states that, when sending an SMS, either the \texttt{body} parameter or the \texttt{media\_url} parameter must be set, but not both at the same time. Similarly, the documentation of the QuickBooks payments API \cite{quickbooks-api} explains that, when creating a credit card, at least one of the parameters \texttt{region} or \texttt{postalCode} must be provided, although both of them are declared as optional.


Current specification languages for RESTful web APIs such as OAS and RAML provide little or no support at all for describing dependencies among input parameters. 
Instead, they just encourage to describe such dependencies as a part of the description of the parameters in natural language, which may result in ambiguous or incomplete descriptions. For example, the Swagger documentation states\footnote{\url{https://swagger.io/docs/specification/describing-parameters/}} ``\emph{OpenAPI 3.0 does not support parameter dependencies and mutually exclusive parameters. (...) What you can do is document the restrictions in the parameter description and define the logic in the 400 Bad Request response}''. The lack of support for dependencies means a strong limitation for current specification languages, since without a formal description of such constraints is hardly possible to interact with the services without human intervention.
For example, it would be extremely difficult to automatically generate test cases for the APIs of Twilio or QuickBooks without an explicit and machine-readable definition of the dependencies mentioned above.
The interest of industry in having support for these types of dependencies is reflected in an open feature request in OAS entitled ``Support interdependencies between query parameters'', created in January 2015 with the message shown below. At the time of writing this paper, the request has received over 260 votes, and it has received 55 comments from 33 participants\footnote{\url{https://github.com/OAI/OpenAPI-Specification/issues/256}}.

\begin{displayquote}
\emph{``It would be great to be able to specify interdependencies between query parameters. In my app, some query parameters become ``required'' only when some other query parameter is present. And when conditionally required parameters are missing when the conditions are met, the API fails. Of course I can have the API reply back that some required parameter is missing, but it would be great to have that built into Swagger.''}
\end{displayquote}

This feature request has fostered an interesting discussion where the participants have proposed different ways of extending OAS to support dependencies among input parameters. However, each approach aims to address a particular type of dependency and thus show a very limited scope. Addressing the problem of modelling and validating input constraints in web APIs should necessarily start by understanding how dependencies emerge in practice. Inspired by this idea, in a previous paper we conducted a thorough study on the presence of inter-parameter dependencies in industrial web APIs~\cite{alberto19icsoc}. For that purpose, we reviewed more than 2.5K operations from 40 real-world RESTful APIs from different application domains. As expected, we found that input dependencies are the norm, rather than the exception, with 85\% of the reviewed APIs having some kind of dependency among their input parameters. More importantly, as the main outcome of our study, we presented a catalogue of seven types of dependencies consistently found in RESTful web APIs. These findings, and specifically the catalogue of dependencies (described in Section~\ref{sec-survey}), serve as the starting point for this work.

In this article, we first present a domain-specific language for the specification of inter-parameter dependencies in web APIs called \emph{Inter-parameter Dependency Language} (\idl). Second, we present an approach for the automated analysis of IDL specifications using constraint programming. In particular, we present a general-purpose mapping showing how to translate an \idl specification into a constraint satisfaction problem (CSP). Then, we present a catalogue of \noperations analysis operations of \idl specifications and show how they can be automated using standard constraint programming reasoning operations. For example, given an \idl specification one may be interested to know if it includes any inconsistencies like parameters that cannot be selected (\emph{dead} parameters) or whether a given call to the API satisfies all the dependencies. Our approach is supported by several tools including an (Eclipse) editor, a parser, an OAS extension (called \emph{IDL4OAS}), and a constraint-programming aided library supporting the automated analysis of \idl specifications. These tools were heavily validated using standard testing techniques resulting in a test suite composed of \ntestcases test cases, which will hopefully facilitate future extensions and alternative implementations. 

The contributions reported in this paper prepare the ground for a new range of specification-driven applications in web APIs. For example, an API gateway supporting the automated analysis of \idl could automatically reject requests violating any dependencies, without even redirecting the call to the corresponding service, saving time and user quota. Also, test case generators supporting the automated analysis of \idl could automatically generate valid test cases (those satisfying all the dependencies among input parameters) rather than using brute force or writing specific input grammars for each API under test. Code generators could benefit from \idl as well. For instance, automatically-generated clients could include built-in assertions to deal with invalid input combinations, preventing input-validation failures. Analogously, interactive API documentations could be enriched with analysis capabilities to detect invalid calls even before invoking the API. The range of new applications is promising.



This paper is structured as follows: Section \ref{sec-survey} presents the catalogue of dependency patterns found in our systematic review of real-world APIs. Section \ref{sec-idl} introduces the syntax of IDL using examples. Section \ref{sec-analysis} describes our approach for the automated analysis of \idl specifications. Our tool suite is presented in Section \ref{sec-tool}. Section \ref{sec-evaluation} describes the evaluation of our approach. Section \ref{sec-threats} describes the possible threats to validity and how these were mitigated. The related work is discussed in Section \ref{sec-related}. Finally, Section \ref{sec-conclusions} concludes the paper and presents future lines of research.


\section{Catalogue of Dependencies}
\label{sec-survey}

The contributions presented in this paper are built on the findings of a previous study by the authors on the presence of inter-parameter dependencies in industrial RESTful Web APIs~\cite{alberto19icsoc}. For the sake of understandability and to make our paper self-contained, we next summarise those results more relevant for this article, and redirect the interested reader to the original paper for further details. 

\begin{figure*}[t]
  \hspace{0.9cm}
   \subfloat[Dependency types.\label{fig:dependencies-per-type}]{%
     \includegraphics[width=0.36\textwidth]{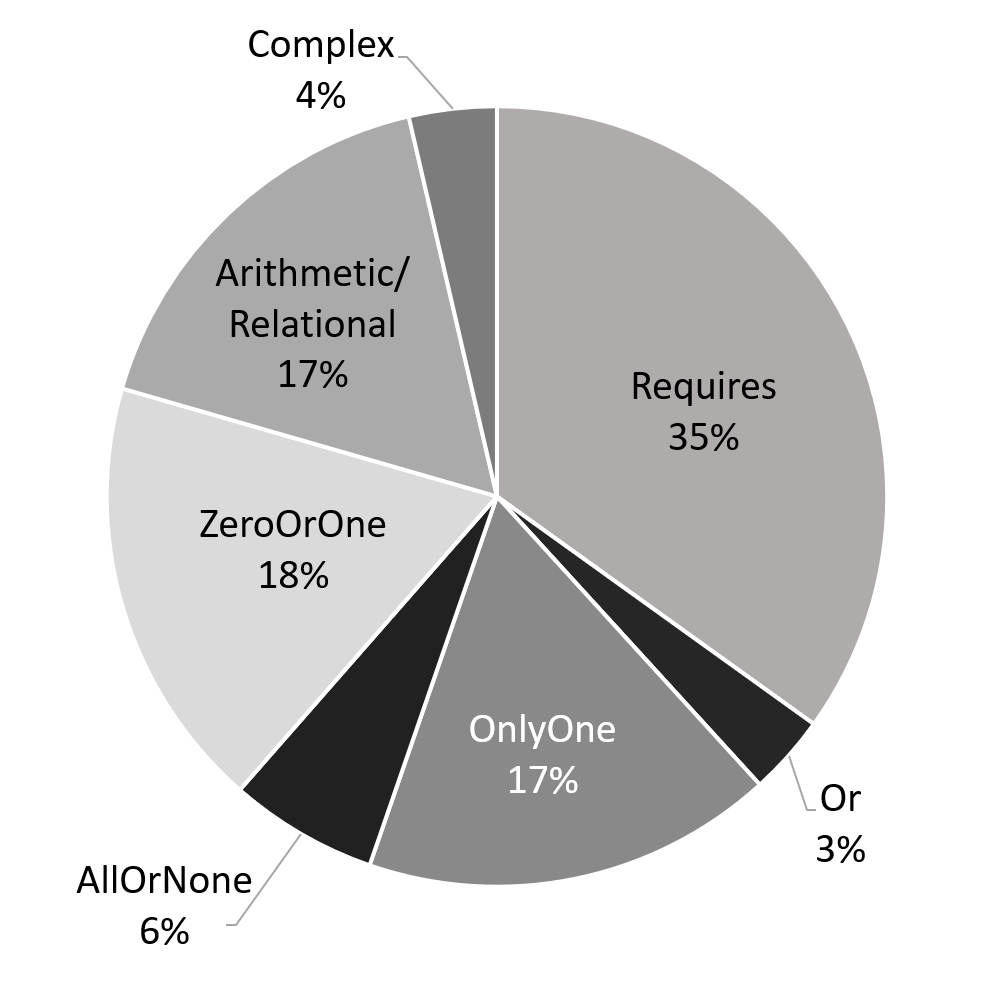}
   }
  \hfill
   \subfloat[Occurrences in APIs \label{fig:dependency-types-per-api-occurrences}]{%
     \includegraphics[width=0.33\textwidth]{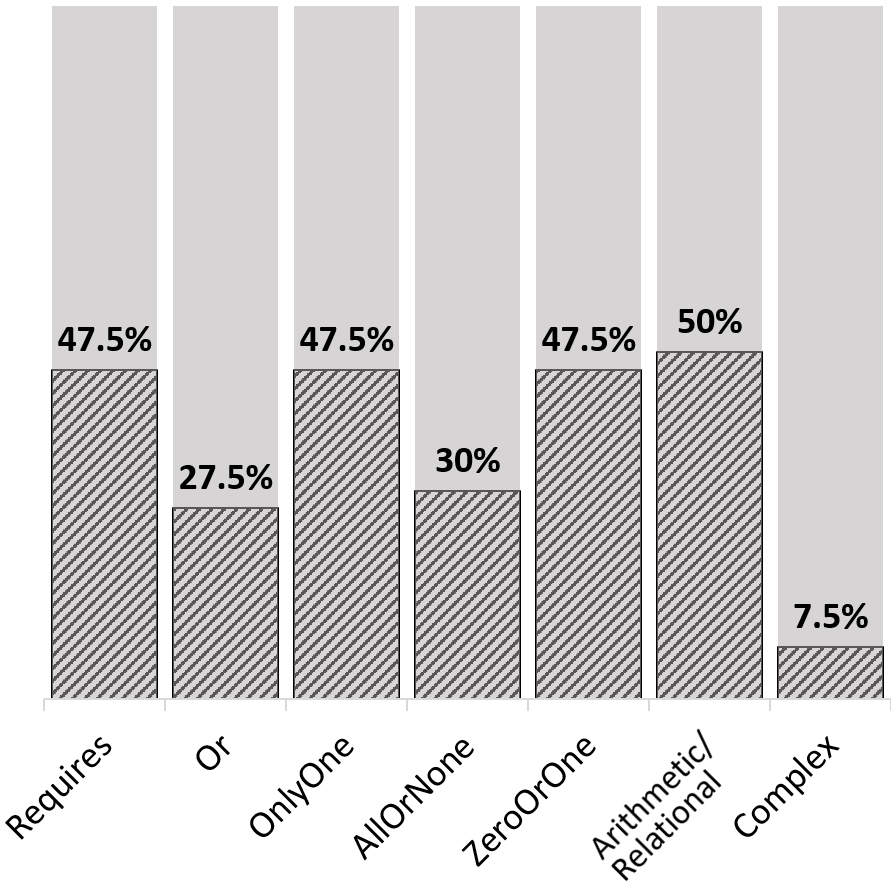}
   }
  \hspace{1.6cm}
   \caption{Distribution of dependencies by type and percentage of APIs.}
   \label{fig:dependency-types}
\end{figure*}

In our previous study, we reviewed more than 2.5K operations from 40 real-world RESTful APIs including popular APIs such as those of YouTube, Google Maps, Amazon S3, and PayPal. The results of the study showed that dependencies are extremely common and pervasive---they appear in 85\% of the APIs under study (34 out of 40) across all application domains and types of operations. Specifically, we identified 633 dependencies among input parameters in 9.7\% of the API operations analysed (248 out of 2,557). The collected data helped us to characterise dependencies identifying their most common shape---dependencies in read operations involving two query parameters---, but also exceptional cases such as dependencies involving up to 10 parameters and dependencies among different types of parameters, e.g., header and body parameters. More importantly, we classified the inter-parameter dependencies identified into seven general types, described below.


Before going in depth into each type of dependency, a number of considerations must be taken into account. First, for the sake of simplicity, dependencies are described using single parameters. However, all dependencies can be generalised to consider groups of parameters using conjunctive and disjunctive connectors. Second, dependencies can affect not only the presence or absence of parameters, but also the values that they can take. 
In what follows, when making reference to a parameter \emph{being present} or \emph{being absent}, it could also mean a parameter \emph{taking a certain value}. 
Finally, when introducing each dependency type we will make reference to Figure~\ref{fig:dependency-types}, which shows the distribution of dependencies by type (Figure~\ref{fig:dependencies-per-type}) and the percentage of subject APIs including occurrences of each dependency type (Figure~\ref{fig:dependency-types-per-api-occurrences}). Next, we describe the seven types of dependencies found in our study, including examples.\\

\noindent \textbf{\dependencyREQUIRES}. The presence of a parameter $p_1$ in an API call requires the presence of another parameter $p_2$, denoted as $p_1 \rightarrow p_2 $. As previously mentioned, $p_1$ and $p_2$ can be generalised to groups of parameters and parameter-value relations, e.g., $a \wedge b=x \rightarrow c \vee d$. Based on our results, this is the most common type of dependency in web APIs, representing 35\% of all the dependencies identified in our study (Figure~\ref{fig:dependencies-per-type}), and being present in 47.5\% of the subject APIs (Figure~\ref{fig:dependency-types-per-api-occurrences}). As an example, in the GitHub API \cite{github-api}, when creating a card in a project, if the parameter \texttt{content\_id} is present, then \texttt{content\_type} becomes required, i.e., \texttt{content\_id} $\rightarrow$ \texttt{content\_type}. Similarly, in the Bing Maps API \cite{bing-maps-api}, when calculating the distance of a set of routes, if the parameter \texttt{startTime} is used, then the parameter \texttt{travelMode} must be set to \texttt{`driving'}, i.e., \texttt{startTime} $\rightarrow$ \texttt{travelMode=driving}.\\



\noindent \textbf{\dependencyOR}. Given a set of parameters $p_1,p_2,\ldots,p_n$, one or more of them must be included in the API call, denoted as $Or(p_1,p_2,\ldots,p_n)$. As illustrated in Figure~\ref{fig:dependency-types}, this type of dependencies represent only 3\% of the dependencies identified in the subject APIs. Interestingly, however, we found that more than one fourth of the APIs (27.5\%) included some occurrence of this type of dependency, which suggests that its use is fairly common in practice. As an example, in the Google Maps Places API \cite{google-maps-api}, when searching for places, both \texttt{query} and \texttt{type} parameters are optional, but at least one of them must be used, i.e., $Or($\texttt{query}, \texttt{type}$)$. Similarly, in the NationBuilder API \cite{nationbuilder-api}, when creating a blog post, it is possible to show different contents on the index page and the full post page by using the parameters \texttt{contentbeforeflip} and \texttt{contentafterflip}, respectively, but at least one of them must be set, i.e., $Or($\texttt{contentbeforeflip}, \texttt{contentafterflip}$)$.\\





\noindent \textbf{\dependencyXOR}. Given a set of parameters $p_1,p_2,\ldots,p_n$, one, and only one of them must be included in the API call, denoted as $OnlyOne(p_1,p_2,\ldots,p_n)$. As observed in Figure~\ref{fig:dependency-types}, this group of dependencies represent 17\% of all the dependencies identified, and they appear in almost half of the APIs under study (47.5\%). Among others, we found that this type of dependency is very common in APIs from the category \emph{media}, where a resource can be identified in multiple ways, e.g., a song can be identified by its name or by its ID, and only one value must be typically provided. For example, in the Last.fm API \cite{lastfm-api}, when getting the information about an artist, this can be identified with two possible parameters, \texttt{artist} or \texttt{mbid}, and only one must be used, i.e., $OnlyOne($\texttt{artist}, \texttt{mbid}$)$. Similarly, in the GeoNames API \cite{geonames-api}, when searching for places, they must be filtered using one, and only one of the parameters \texttt{q}, \texttt{name} and \texttt{name\_equals}, i.e., $OnlyOne($\texttt{q}, \texttt{name}, \texttt{name\_equals}$)$.\\


\begin{figure}[t]
  \centering
  \includegraphics[width=0.48\textwidth, cfbox=darkgray 0.1pt 0.1pt]{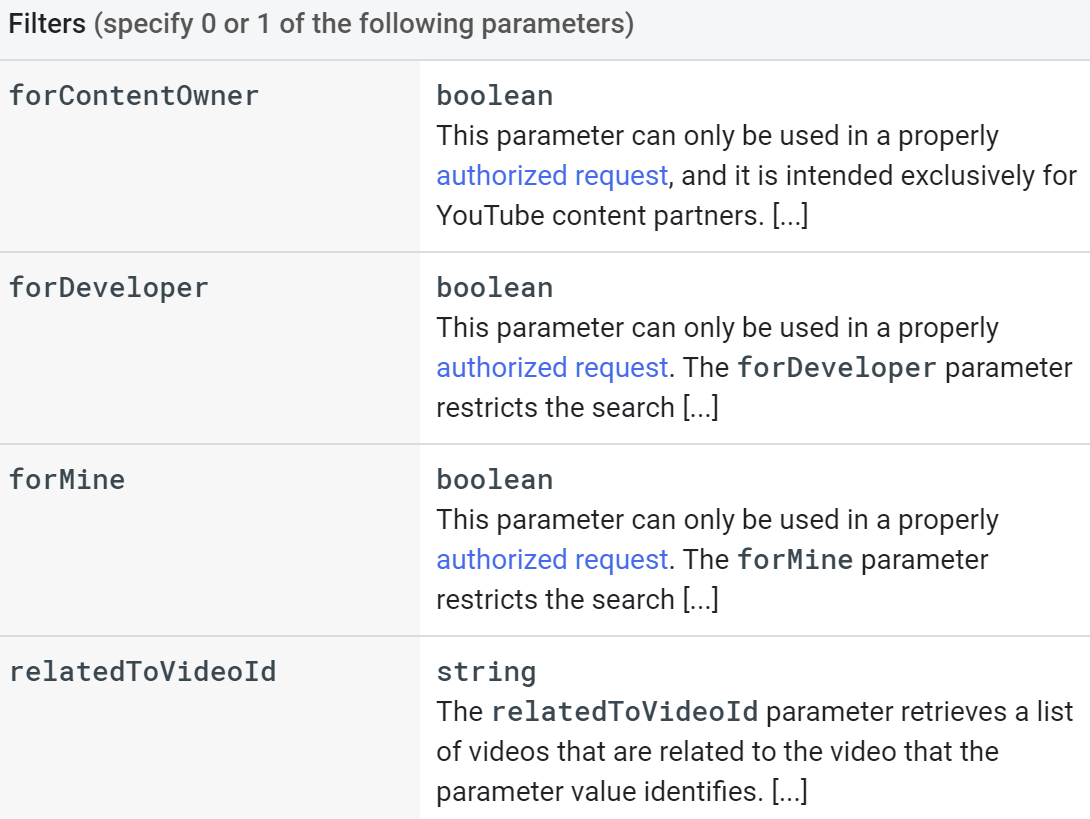}
  \caption{\emph{ZeroOrOne} dependency in the YouTube API.} \label{fig:youtube-dep}
\end{figure}



\noindent \textbf{\dependencyXNOR}. Given a set of parameters $p_1,p_2,\ldots,p_n$, either all of them are provided or none of them, denoted as $AllOrNone(p_1,p_2,\ldots,p_n)$. Very similarly to the \emph{\dependencyOR} dependency type, only 6\% of the dependencies found belong to this category, nonetheless, they are present in about one third of the APIs under study (30\%). In the Yelp API \cite{yelp-api}, for example, when searching for events, the location can optionally be specified with two parameters, \texttt{latitude} and \texttt{longitude}, and they must be used together, i.e., $AllOrNone($\texttt{latitude}, \texttt{longitude}$)$.
In the Bing Web Search API \cite{bing-api}, when using the \texttt{Accept-Language} header, the \texttt{cc} (country code) query parameter must be specified too, and vice versa, i.e., $AllOrNone($\texttt{Accept-Language}, \texttt{cc}$)$.\\





\noindent \textbf{\dependencyNAND}. Given a set of parameters $p_1,p_2,\ldots,p_n$, zero or one can be present in the API call, denoted as $ZeroOrOne(p_1,p_2,\ldots,p_n)$. Figure \ref{fig:dependency-types} reveals that this dependency type is common both in terms of the number of occurrences (18\% of the total) and the number of APIs including it (47.5\%). As an example, in the Flickr API \cite{flickr-api}, when fetching photos from the user's contacts, either one or multiple photos can be obtained by using the parameters \texttt{single\_photo} or \texttt{count}, respectively, but they cannot be used together (if none is used, all photos are returned), i.e., $ZeroOrOne($\texttt{single\_photo}, \texttt{count}$)$.
Another example is found in the search operation of the YouTube API \cite{youtube-api}, where it is possible to filter results with four optional but mutually exclusive parameters, as depicted in Figure \ref{fig:youtube-dep}. This dependency can be expressed as $ZeroOrOne($\texttt{forContentOwner}, \texttt{forDeveloper}, \texttt{forMine}, \texttt{relatedToVideoId}$)$.\\

\noindent \textbf{\dependencyARITHMETIC}. Given a set of parameters $p_1,p_2,\ldots,p_n$, they are related by means of arithmetic and/or relational constraints, e.g., $p_1 + p_2 > p_3$. As shown in Figure \ref{fig:dependency-types}, this type of dependency is the most recurrent across the subject APIs, being present in half of them. Moreover, 17\% of the dependencies found are of this type. As an example, in the Twitter API \cite{twitter-api}, when searching for tweets, the \texttt{max\_id} parameter must be greater than or equal to the \texttt{since\_id} parameter, otherwise no tweets will be returned, i.e., \texttt{max\_id} $>=$ \texttt{since\_id}. In the payments API Forte \cite{forte-api}, when creating a merchant application, this can be owned by several businesses, in which case the sum of the percentages cannot be greater than 100, i.e., \texttt{owner.percentage} $+$ \texttt{owner2.percentage} $+$ \texttt{owner3.percentage} $+$ \texttt{owner4.percentage} $<=$ \texttt{100}.\\



\begin{figure}[t]
  \centering
  \includegraphics[width=0.45\textwidth, cfbox=darkgray 0.1pt 0.1pt]{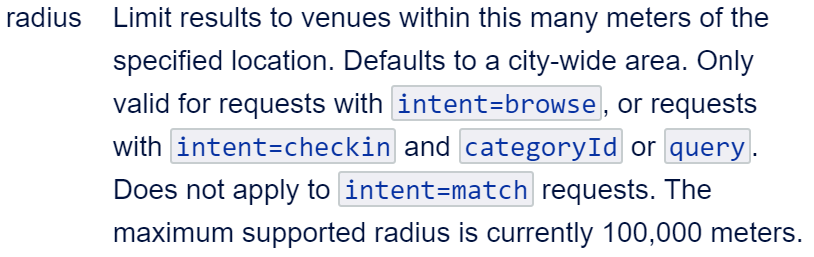}
  \caption{\emph{Complex} dependency present in the \texttt{GET /venues/search} operation of the Foursquare API.} \label{fig:foursquare-dep}
\end{figure}

\noindent \textbf{\dependencyCOMPLEX}. These dependencies involve two or more of the types of constraints previously presented. Based on our results, they are typically formed by a combination of \emph{\dependencyREQUIRES} and \emph{\dependencyXOR} dependencies. As illustrated in Figure \ref{fig:dependency-types}, we found 4\% of complex dependencies, being present in 7.5\% of the subject APIs. For example, in the Stripe API \cite{stripe-api}, when retrieving the information about an upcoming invoice, if \texttt{subscription\_trial\_end} is present, then one of \texttt{subscription\_items} or \texttt{subscription} must be set too, i.e., \texttt{subscription\_trial\_end} $\rightarrow OnlyOne($\texttt{subscription\_items}, \texttt{subscription}$)$.
Figure \ref{fig:foursquare-dep} shows an extract of the documentation of the search operation in the Foursquare API \cite{foursquare-api}. As illustrated, if \texttt{radius} is used, then either \texttt{intent} is set to \texttt{`browse'} or \texttt{intent} is set to \texttt{`checkin'} and \texttt{categoryId} or \texttt{query} are present too, i.e., \texttt{radius} $\rightarrow OnlyOne($\texttt{intent=browse}, \texttt{intent=checkin} $\wedge$ (\texttt{categoryId} $\vee$ \texttt{query}$))$.

\section{Inter-parameter Dependency Language}
\label{sec-idl}

In this section, we present \emph{Inter-parameter Dependency Language} (\idl), a textual domain-specific language for the specification of dependencies among input parameters in web APIs. Specifically, \idl is designed to express the seven types of inter-parameter dependencies identified in our study on real-world APIs, and described in the previous section. For the design of the language, we took inspiration from the input format of the combinatorial testing tool Pairwise Independent Combinatorial Testing (PICT)~\cite{pict}, by Microsoft, where constraints among input parameters can be defined using invariants, conditional definitions (if/then/else), logical operators and relational operators.

It is worth mentioning that \idl focuses on the definition of dependencies among parameters, but not in the definition of the parameters themselves. This is because \idl is specifically designed to be easily integrated into API specification languages such as OAS or RAML, where parameters are specified in different ways. Thus, in what follows, we simply assume that each parameter has a name and a domain. 

A simplified version of the grammar of the language is provided in Listing~\ref{list:idl-grammar}---the full version is available as a part of the implementation of IDL \cite{idl} and on the supplemental material provided with this article \cite{landing-page}.

\lstset{
  language=Grammar,
  basicstyle=\scriptsize\ttfamily,
  numbers=left,
  numberstyle=\tiny\color{codegray},
  numbersep=9pt,
  xleftmargin=15pt,
  frame=single,
  framesep=3pt,
  rulecolor=\color{backcolour}
}
\begin{lstlisting}[style=listingtop, caption={Simplified grammar of IDL.}, label={list:idl-grammar}]
Model:
  Dependency*;
Dependency:
    RelationalDependency | ArithmeticDependency |
    ConditionalDependency | PredefinedDependency;
RelationalDependency:
  Param RelationalOperator Param;
ArithmeticDependency:
  Operation RelationalOperator DOUBLE;
Operation:
  Param OperationContinuation |
  '(' Operation ')' OperationContinuation?;
OperationContinuation:
  ArithmeticOperator (Param | Operation);
ConditionalDependency:
  'IF' Predicate 'THEN' Predicate;
Predicate:
    Clause ClauseContinuation?;
Clause:
    (Term | RelationalDependency | ArithmeticDependency
    | PredefinedDependency) | 'NOT'? '(' Predicate ')';
Term:
    'NOT'? (Param | ParamValueRelation);
Param:
  ID | '[' ID ']';
ParamValueRelation:
  Param '==' STRING('|'STRING)* |
  Param 'LIKE' PATTERN_STRING | Param '==' BOOLEAN |
  Param RelationalOperator DOUBLE;
ClauseContinuation:
  ('AND' | 'OR') Predicate;
PredefinedDependency:
  'NOT'? ('Or' | 'OnlyOne' | 'AllOrNone' |
  'ZeroOrOne') '(' Clause (',' Clause)+ ')';
RelationalOperator:
  '<' | '>' | '<=' | '>=' | '==' | '!=';
ArithmeticOperator:
  '+' | '-' | '*' | '/';
\end{lstlisting}

The key elements of the language are terms and predicates. Both of them can evaluate to true or false. A \emph{term} is an atomic element of the language and can be represented by: (1) a parameter's name (e.g., \texttt{p1}) being evaluated as true if the parameter is set (regardless of the value), or false otherwise; or (2) a parameter-value relation, evaluated as true if the parameter is selected and satisfies the relation. 
This relation can be defined using standard relational operators (e.g., \texttt{p1>=100}) or a wild card match---using the operator LIKE---if the parameter is a string, with '*' meaning zero or more characters and '?' meaning one character (e.g., \texttt{p3 LIKE `test\_*'}).
A \emph{predicate} is a combination of one or more terms and dependencies joined by the logical operators NOT, AND, and OR. 
Parentheses are allowed in order to specify the operator priority. In what follows, we describe the \idl notation of each type of dependency.


\lstset{
  language=IDL,
  numbers=left,
  numberstyle=\tiny\color{codegray},
  numbersep=9pt,
  xleftmargin=15pt,
  xrightmargin=3.5pt,
  frame=single,
  framesep=3pt,
}

\vspace{0.35cm}



\noindent \textbf{\dependencyREQUIRES}. This type of dependency is expressed as ``\texttt{\footnotesize{IF predicate THEN predicate;}}'', where the first predicate is the \emph{condition} and the second is the \emph{consequence}. The following listing shows two examples. Dependency in line 2, for instance, indicates that invocations including the parameters \texttt{p1} and \texttt{p2} should not include \texttt{p3} nor \texttt{p4}, otherwise the call would be invalid.

\begin{lstlisting}
IF p1 THEN p2=='A';
IF p1 AND p2 THEN NOT (p3 OR p4);
\end{lstlisting}

\vspace{0.35cm}




\noindent \textbf{\dependencyOR}. This type of dependency is expressed using the keyword ``Or'' followed by a list of two or more predicates placed inside parentheses: ``\texttt{\footnotesize{Or(predicate, predicate [, ...]);}}''. The dependency is satisfied if at least one of the predicates evaluates to true. Two examples follow. Dependency in line 1, for instance, specifies that valid invocations should include at least one of the parameters \texttt{p1}, \texttt{p2} or \texttt{p3}.

\begin{lstlisting}
Or(p1, p2, p3);
Or(p1, p3 AND p5, p6=='B');
\end{lstlisting}



\vspace{0.35cm}


\noindent \textbf{\dependencyXOR}. These dependencies are specified using the keyword ``OnlyOne'' followed by a list of two or more predicates placed inside parentheses: ``\texttt{\footnotesize{OnlyOne(predicate, predicate [, ...]);}}''. The dependency is satisfied if one, and only one of the predicates evaluates to true. Examples of this dependency are shown below. The dependency in line 1, for example, indicates that valid invocations should include either the parameter \texttt{p1} or the parameter \texttt{p2} with value \texttt{`B'}, but not both at the same time. 

\begin{lstlisting}
OnlyOne(p1, p2=='B');
OnlyOne(p1 OR p2, p3 AND (p4 OR p5));
\end{lstlisting}

\vspace{0.35cm}


\noindent \textbf{\dependencyXNOR}. This type of dependency is specified using the keyword ``AllOrNone'' followed by a list of two or more predicates placed inside parentheses: ``\texttt{\footnotesize{AllOrNone(predicate, predicate [, ...]);}}''. The dependency is satisfied if either all the predicates evaluate to true, or all of them evaluate to false. The dependency in line 1 below, for instance, indicates that valid calls are those including either the parameter \texttt{p1} and \texttt{p2} with value \texttt{true}, or conversely, those not including the parameter \texttt{p1} and not including \texttt{p2} with value \texttt{true}.

\begin{lstlisting}
AllOrNone(p1, p2==true);
AllOrNone(p1 AND p2, p3 LIKE 'test_*' OR p4<10);
\end{lstlisting}

\vspace{0.35cm}


\noindent \textbf{\dependencyNAND}. These dependencies are specified using the keyword ``ZeroOrOne'' followed by a list of two or more predicates placed inside parentheses: ``\texttt{\footnotesize{ZeroOrOne(predicate, predicate [, ...]);}}''. The dependency is satisfied if none or at most one of the predicates evaluates to true. Two examples follow. Line 2, for instance, specifies that valid invocations must meet zero or one (but not both) of the two conditions between parentheses: (1) including the parameter \texttt{p1}, or (2) including the parameter \texttt{p2} with a value less than or equal to 100. 

\begin{lstlisting}
ZeroOrOne(p1, p2, p3, p4);
ZeroOrOne(p1, p2<=100);
\end{lstlisting}

\vspace{0.35cm}


\noindent \textbf{\dependencyARITHMETIC}. Relational dependencies are specified as pairs of parameters joined by any of the following relational operators: ==, !=, <=, <, >= or > (see examples in lines 1 and 2 below). Arithmetic dependencies relate two or more parameters using the operators +, - , *, / followed by a final comparison using a relational operator. 
Lines 3 and 4 of the following listing show examples of arithmetic dependencies.

\begin{lstlisting}
p1 < p2;
p1 != p2;
p1 + p2 - p3 * p4 == 100;
p1 * p2 / ((p3 - p4) * p5) < 176.89;
\end{lstlisting}

\vspace{0.35cm}


\noindent \textbf{\dependencyCOMPLEX}. These dependencies are specified as a combination of the previous ones since, as previously mentioned, predicates can contain other dependencies. As an exception to this rule, predicates cannot include \emph{Requires} dependencies to avoid over-complicated specifications (such dependencies can be expressed in simpler ways). The following listing shows some examples of complex dependencies. Dependency in line 1 combines four different types of dependencies: \emph{Requires}, \emph{ZeroOrOne}, \emph{OnlyOne} and \emph{Relational}.


\begin{lstlisting}
IF p1 THEN ZeroOrOne(p2, OnlyOne(p3, p4>p5));
AllOrNone(p1+p2<100, Or(p3=='A', Or(p4, p5>p6));
\end{lstlisting}


\vspace{0.35cm}

It is worth making a few general clarifications about the language regarding dependencies \emph{Or}, \emph{OnlyOne}, \emph{AllOrNone} and \emph{ZeroOrOne}. These are not strictly necessary, as they could be translated to several \emph{Requires} dependencies. However, they are provided as syntactic sugar to make specifications succinct and self-explanatory. An example is given in the following \idl excerpt (lines 1-3). Secondly, they cannot contain negated elements within their parentheses, since such constraints can be expressed in simpler ways (lines 5-6). Finally, they can optionally be preceded by the keyword ``NOT'' to negate the meaning of the constraint (see line 8 below for an example).

\begin{lstlisting}
AllOrNone(p1, p2);   // Equivalent to 1) and 2):
IF p1 THEN p2;       // 1)
IF p2 THEN p1;       // 2)

Or(p1, NOT p2);      // Invalid dependency
IF p2 THEN p1;       // Equivalent to line 5

NOT OnlyOne(p1, p2); // Valid negated dependency
\end{lstlisting}



Listing~\ref{list:googlemaps-idl} depicts the \idl specification of the Google Maps Places API \cite{google-maps-api}. It comprises seven operations, four of which have dependencies. The API has eight dependencies in total, including six out of the seven types of dependencies supported in \idl (all of them except the complex ones), namely:

\begin{itemize}
    \item Line 2: If the parameter \texttt{radius} is used, then \texttt{rankby} cannot be set to \texttt{`distance'}, and vice versa.
    \item Line 3: If the parameter \texttt{rankby} is set to \texttt{`distance'}, then at least one of the following parameters must be present: \texttt{keyword}, \texttt{name} or \texttt{type}.
    \item Line 4: The parameter \texttt{maxprice} must be greater than or equal to \texttt{minprice}.
    \item Line 7: Either both \texttt{location} and \texttt{radius} are used, or none of them.
    \item Line 8: \texttt{query} and \texttt{type} are both optional parameters, but at least one of them must be used.
    \item Line 9: Equal to line 4.
    \item Line 12: One, and only one of the parameters \texttt{maxheight} and \texttt{maxwidth} must be used.
    \item Line 15: If the parameter \texttt{strictbounds} is used, then both \texttt{location} and \texttt{radius} must be used as well.
\end{itemize}


\lstset{
  numbers=left,
  numberstyle=\tiny\color{codegray},
  numbersep=9pt,
  xleftmargin=15pt,
  xrightmargin=3.5pt,
  frame=single,
  framesep=3pt,
}

\begin{lstlisting}[caption={\idl specification of Google Maps Places API.}, label={list:googlemaps-idl}]
// Operation: Search for places within specified area:
ZeroOrOne(radius, rankby=='distance');
IF rankby=='distance' THEN keyword OR name OR type;
maxprice >= minprice;

// Operation: Query information about places:
AllOrNone(location, radius);
Or(query, type);
maxprice >= minprice;

// Operation: Get photo of place:
OnlyOne(maxheight, maxwidth);

// Operation: Automcomplete place name:
IF strictbounds THEN location AND radius;
\end{lstlisting}

\section{Automated Analysis}
\label{sec-analysis}

The analysis of \idl deals with the extraction of information from \idl specifications. For example, given an \idl specification, we might be interested to know whether it contains errors (e.g., inconsistent dependencies) or whether a given API call is valid, i.e., it meets all the constraints defined in the specification. Performing these analyses manually is hardly possible in practice.

In what follows, we present our approach for the automated analysis of \idl specifications using constraint programming. In particular, we first present the formal semantics of \idl by explaining how \idl specifications can be mapped to a constraint satisfaction problem (CSP). Then, we present a catalogue of \noperations analysis operations of \idl specifications and show how they can be automated using standard constraint programming reasoning operations. 

\subsection{Formal Semantics of IDL}
\label{sec-idlSemantics}

The primary objective of formalising IDL is to establish a sound basis for the automated support. Following the formalisation principles defined by Hofstede et al. \cite{Hofstede98}, we follow a transformational style by translating IDL specifications to a target domain suitable for the automated analysis (\emph{Primary Goal Principle}). Specifically, we propose translating IDL specifications to a CSP that can be then analysed using state-of-the-art constraint programming tools. A similar approach was followed by the authors to automate the analysis of feature models~\cite{benavides10} and service level agreements~\cite{Muller13TSC,Muller18TSC}.

A CSP is defined as a 3-tuple $(V,D,C)$ composed of a set of variables $V$, their domains $D$ and a number of constraints $C$. A solution for a CSP is an assignment of values to the variables in $V$ from their domains in $D$ so that all the constraints in $C$ are satisfied. 


\setlength{\extrarowheight}{3pt}
\begin{table*}[t]
\footnotesize
\begin{center}\begin{tabular}{|p{0.035\textwidth}|p{0.25\textwidth}|p{0.61\textwidth}|}
\hline
\rowcolor[rgb]{0.8,0.8,0.8}
\multicolumn{3}{|c|}{\textbf{Mapping from IDL to CSP}}\\
\rowcolor[rgb]{0.9,0.9,0.9}
\multicolumn{2}{|c|}{API Parameters} & \multicolumn{1}{|c|}{CSP Mapping}\\\hline
& \vspace{-0.5cm}[Parameters] $P$ & $\forall p_i \in P, \left \{ \begin{array}{l} V \leftarrow V \cup p_{i} \cup p_{i}Set \\ 
D \leftarrow D \cup domain(p_{i}) \cup Boolean \\
C \leftarrow C \cup p_{i}Set==true$ (if $p_{i}$ is required) $\end{array}\right.$ \\ \hline
\rowcolor[rgb]{0.9,0.9,0.9}
\multicolumn{2}{|c|}{IDL Element} & \multicolumn{1}{|c|}{CSP Mapping}\\\hline
\multirow{5}{1.5cm}{
\hspace{-0.3cm}
\rotatebox{90}
    {\begin{tabular}{@{}c}
            Terms:\\
            map(T)
        \end{tabular}}    
} & [Parameter] & \\ 
& \vspace{-0.4cm}\hspace{0.2cm}$p_{i}$ & \vspace{-0.4cm} $C \leftarrow C \cup \{p_{i}Set == true\}$ \\\cline{2-3}

& [Parameter-Value Relation] &  \\
& \hspace{0.2cm}$p_i \; relOp^{\star} \; v$ & \vspace{-0.4cm} $C \leftarrow C \cup \{p_i \; relOp \; v \wedge \; p_{i}Set == true\}$ \\\cline{1-3}
\multirow{7}{1.2cm}{
\hspace{-0.3cm}
\rotatebox{90}
    {\begin{tabular}{@{}c}
            Predicates:\\
            map(P)
        \end{tabular}}   
} & [Term] \hspace{0.2cm}$T$ & $map(T)$\\  \cline{2-3}
& [Dependency] \hspace{0.2cm}$D$ & $map(D)$ \\ \cline{2-3}

& [Term AND Predicate] & \\ 
& \hspace{0.2cm}$T$ AND $P$ & \vspace{-0.4cm}$C \leftarrow C \cup map(T) \wedge map(P)$ \\ \cline{2-3}

& [Term OR Predicate] & \\
& \hspace{0.2cm}$T$ OR $P$  & \vspace{-0.4cm} $C \leftarrow C \cup map(T) \vee map(P)$ \\ \cline{2-3} 

& [NOT Predicate] & \\ 
& \hspace{0.2cm}NOT $P$  & \vspace{-0.4cm} $C \leftarrow C \cup \neg map(P)$ \\ \cline{1-3}

\multirow{13}{1.2cm}{
\hspace{-0.3cm}
\rotatebox{90}
    {\begin{tabular}{@{}c}
            Dependencies:\\
            map(D)
        \end{tabular}}    
} & [Requires] & \\
& \hspace{0.2cm}IF $P_{i}$ THEN $P_{j}$ &  \vspace{-0.4cm} $C \leftarrow C \cup map(P_{i}) \implies map(P_{j})$ \\ \cline{2-3}

& [Or] &  \\
& \hspace{0.2cm}$Or(P_{1}, ..., P_{n})$   & \vspace{-0.4cm} $C \leftarrow C \cup \bigvee_{i=1}^{n} map(P_{i})$ \\ \cline{2-3}

& [OnlyOne] & \\
& \hspace{0.2cm}$OnlyOne(P_{1}, ..., P_{n})$   &  \vspace{-0.4cm}$C \leftarrow C \cup  \{\forall_{i=1}^{n}, \forall_{j=1}^{n} | i \neq j, map(P_{i}) \implies \neg map(P_{j})\}$ \\ \cline{2-3}

& [AllOrNone] &  \\ 
& \hspace{0.2cm}$AllOrNone(P_{1}, ..., P_{n})$   & \vspace{-0.4cm}$C \leftarrow C \cup \forall_{i=1}^{n}, \forall_{j=1}^{n} | i \neq j, \{map(P_{i}) \implies map(P_{j})\} \wedge \{\neg map(P_{i}) \implies \neg map(P_{j})\}$\\ \cline{2-3}

& [ZeroOrOne] & \\
& \hspace{0.2cm}$ZeroOrOne(P_{1}, ..., P_{n})$ & \vspace{-0.4cm}$C \leftarrow C \cup \{map(OnlyOne(P_{1}, ..., P_{n}))\} \vee \{\bigwedge_{i=1}^{n} \neg map(P_{i})\} $ \\ \cline{2-3}

& [Relational Dependency] &  \\ 
& \vspace{-0.4cm}\hspace{0.2cm}$p_i \; relOp^{\star} \; p_j$ & \vspace{-0.4cm}$C \leftarrow C \cup \{(p_{i}Set==true \wedge p_{j}Set==true) \implies p_i \; relOp \; p_j\}$ \\\cline{2-3}

& [Arithmetic Dependency] &  \\
& \hspace{0.2cm}$p_i \; arOp^{\diamond} \; p_j \; arOp ... \; p_n \; relOp^{\star} \; v$ & $C \leftarrow C \cup \{(p_{i}Set==true \wedge p_{j}Set==true \wedge \; ... \; p_{n}Set==true)$\\
& & \hspace{0.68cm}$\implies (p_i \; arOp \; p_j \; arOp \; ... \; p_n \; relOp \; v)\}$ \\
\hline
\multicolumn{3}{l}{\vspace{-0.1cm}\scriptsize{$\star \; relOp = \{<|==|\neq|\geq|\leq|>\}$}}\\
\multicolumn{3}{l}{\vspace{-0.1cm}\scriptsize{$\diamond \; arOp = \{+|-|*| \div\}$}}
\end{tabular}
\caption{IDL to CSP mapping}
\label{tab:MappingIDL}  
\end{center}
\vspace{-0.5cm}
\end{table*}

Table~\ref{tab:MappingIDL} describes the mapping from \idl to CSP. The first row of the table depicts how each input parameter is mapped to CSP variables, domains and constraints. Recall that both the name and domain of each parameter should be taken from the API specification (c.f. $p_{i}$ and $domain()$ function). For each parameter, two CSP variables are created: (1) one representing the parameter itself (c.f. $p_{i}$), and (2) a Boolean variable to express whether the parameter is set or not (c.f. $p_{i}Set$). Optionally, we may also get information from the specification about whether each parameter is required (mandatory) or not. If a parameter $p_{i}$ is required (i.e., it must be present in all API calls), the constraint $p_{i}Set==true$ is added to the set of constraints $C$. The second and third rows of the mapping in Table~\ref{tab:MappingIDL} express how the terms are mapped to a CSP.
Every time a parameter is found in a predicate, it must be checked whether the parameter is present in the API request. If so, it will evaluate to true, otherwise it will evaluate to false (c.f. $p_{i}Set==true$ from the second row of the table). In the case of parameters having a relational condition with a value, it must also be checked that the parameter satisfies such condition (c.f. third row of the table). Finally, predicates and dependencies are defined recursively using the function $map(E)$, where $E$ is either a term, a predicate or a dependency. Exceptionally, relational and arithmetic dependencies are only evaluated if all the involved parameters are present in the API request (c.f. last two rows in Table~\ref{tab:MappingIDL}).

\begin{lstlisting}[caption={CSP of \emph{Search} operation in Listing \ref{list:googlemaps-idl}.}, label={list:csp},language=MAP,mathescape=true,escapeinside={(*@}{@*)}]
V = $\{$ radius, radiusSet, rankby, rankbySet, keyword,
      keywordSet, name, nameSet, type, typeSet,
      maxprice, maxpriceSet, minprice, minpriceSet    $\}$
      
D = $\{$ int, Boolean, string, Boolean, string, Boolean,
      string, Boolean, string,  Boolean, int, Boolean,
      int, Boolean                                    $\}$
      
C = $\{$//ZeroOrOne(radius, rankby=='distance');
    ((radiusSet==true $\implies$$\neg$(rankbySet==true AND
    rankby==distance) AND ((rankbySet==true AND
    rankby==distance) $\implies$$\neg$radiusSet==true)) OR
    (($\neg$radiusSet==true) AND $\neg$(rankbySet==true
    AND rankby==distance))) AND
    //IF rankby=='distance' THEN keyword OR name OR
    //   type;
    ((rankbySet==true AND rankby==distance)$\implies$
    ((keywordSet==true) OR (nameSet==true) OR
    (typeSet==true))) AND
    //maxprice >= minprice;
    (((maxpriceSet==true) AND (minpriceSet==true)) $\implies$
    (maxprice $\geq$ minprice))                             $\}$ 
\end{lstlisting}

As an example, Listing \ref{list:csp} shows the resulting CSP obtained as a result of applying the proposed mapping to the IDL specification of the \emph{Search} operation in the Google Maps Places API, specified in Listing \ref{list:googlemaps-idl} (lines 1-4). Analogously, Listings \ref{list:cspQuery} and \ref{list:cspGet} depict the CSP constraints derived from the \emph{Query} and \emph{Get} operations in Listing \ref{list:googlemaps-idl}, respectively (lines 6-9 and 11-12).


\begin{lstlisting}[caption={Constraints of \emph{Query} operation in Listing \ref{list:googlemaps-idl}.}, label={list:cspQuery},language=MAP,mathescape=true,escapeinside={(*@}{@*)}]
C = $\{$//AllOrNone(location, radius);
    ((locationSet==true $\implies$ radiusSet==true) AND 
    (radiusSet==true $\implies$ locationSet==true) AND 
    ($\neg$locationSet==true $\implies$$\neg$radiusSet==true) AND 
    ($\neg$radiusSet==true$\implies$$\neg$locationSet==true)) AND
    //Or(query, type);
    (querySet==true OR typeSet==true) AND
    //maxprice >= minprice;
    (((maxpriceSet==true) AND (minpriceSet==true)) $\implies$
    (maxprice $\geq$ minprice))                             $\}$
\end{lstlisting}

\begin{lstlisting}[caption={Constraints of \emph{Get} operation in Listing \ref{list:googlemaps-idl}.}, label={list:cspGet},language=MAP,mathescape=true,escapeinside={(*@}{@*)}]
C = $\{$//OnlyOne(maxheight, maxwidth);
   ((maxheightSet==true$\implies$$\neg$maxwidthSet==true) AND
   (maxwidthSet==true$ \implies$$\neg$maxheightSet==true))             $\}$ 
\end{lstlisting}

\subsection{Analysis Operations}
\label{sec-analysisOperations}

In this section, we propose a catalogue of \noperations analysis operations on IDL specifications. These operations leverage the formal description of the dependencies using IDL to extract helpful information such as identifying inconsistencies or checking whether an API call is valid or not. Analogous analysis operations have been defined in the context of the automated analysis of feature models~\cite{benavides10} and service level agreements~\cite{Muller13TSC,Muller18TSC}. We may remark that it is not our intention to propose an exhaustive set of analysis operations as that would exceed the scope of this article.



For the description of the operations in CSP, we will refer to the input IDL specification \texttt{IDL} and the list of parameters from the API specification \texttt{P}. Additionally, we will use the following auxiliary operations:

\begin{itemize}
    \item \texttt{map(IDL,P)}. This operation translates an input IDL specification \texttt{IDL} and the list of parameters \texttt{P} from the API specification to a CSP following the mapping described in Section~\ref{sec-idlSemantics}.
    
    \item \texttt{solve(CSP)}. This standard CSP-based operation returns a random solution for the input \texttt{CSP} (if any).
    
    \item \texttt{solveAll(CSP)}. This standard CSP-based operation returns all the solutions of the input \texttt{CSP} (if any).
    
    \item \texttt{filter(CSP,L)}. This operation takes as input a \texttt{CSP} and a list \texttt{L} of pairs variable-value to be set, $\{\{p_1, v_1\},\{p_2,v_2\}, \ldots, \{p_n,v_n\}\}$, and returns the input \texttt{CSP} with additional constraints setting each variable in \texttt{L}, $p_i$, to its corresponding value $v_i$, i.e., $C \leftarrow C \cup \{p_i = v_i\}$ .
\end{itemize}


In what follows, for each operation, we provide a name, a description, an example, and an explanation of how it is mapped to a CSP. 

\lstset{
  numbers=left,
  numberstyle=\tiny\color{codegray},
  numbersep=9pt,
  xleftmargin=15pt,
  xrightmargin=3.5pt,
  frame=single,
  framesep=3pt,
}

\vspace{0.2cm}

\noindent \textbf{\consistency}. This operation receives as input an \idl specification and the list of parameters included on it, and returns a Boolean indicating whether the specification is consistent or not. An \idl specification is \emph{consistent} if there exists at least one request satisfying all the dependencies of the specification. Inconsistent specifications are obviously undesired and therefore automating their detection can be very helpful. This operation can be translated to a CSP as follows:


\begin{align*}
\mathrm{isConsistentIDL(IDL,P) \iff \mathrm{solve(\mathrm{map}(IDL,P))} \neq \emptyset}
\end{align*}

\noindent \textbf{\deadParameter}. This operation takes as input an IDL specification, the list of parameters in the specification, and the name of a parameter, and it returns a Boolean indicating whether the parameter is dead or not. A parameter is \emph{dead} if it cannot be included in any valid call to the service. Dead parameters are caused by inconsistencies in the specification or the design of the service. They may be hard to detect when the inconsistency is caused by several inter-related dependencies. For example, in the following \idl specification, the parameter \texttt{p1} is dead since both constraints cannot be satisfied at the same time.


\begin{lstlisting}
IF p1 THEN p2;
OnlyOne(p1, p2);
\end{lstlisting}

Given an input parameter \texttt{p}, this operation can be automated by setting the CSP variable representing the presence of \texttt{p} to true (${pSet=true}$) and checking whether the problem has at least one solution. If there is no solutions, it means that \texttt{p} is dead, namely:



 \begin{align*}
 \mathrm{isDead}&\mathrm{Parameter(IDL,P,p)} \iff \\
 & \mathrm{solve(filter(\mathrm{map}(IDL,P),\{\{pSet,true\}\})) = \emptyset}
 \end{align*}
 


\noindent \textbf{\falseOptional}. This operation assumes that the specification of each parameter indicates, as in OAS, whether the parameter is required (i.e., it must be included in every service request) or optional. This operation takes as input an \idl specification, its parameters, and the name of a parameter specified as optional, and returns a Boolean indicating whether the parameter is false optional or not. A parameter is \emph{false optional} if it is required (i.e., it must be included in all API calls to satisfy inter-parameter dependencies) despite being defined as optional. False optional parameters should be avoided since they give the user a wrong idea of the domain. For example, suppose that the parameter \texttt{p1} is defined as mandatory (e.g., \texttt{``required'': true} in OAS) and \texttt{p2} is declared as optional (\texttt{``required'': false}). The constraint ``\texttt{IF p1 THEN p2}'' in IDL would make \texttt{p2} a false optional parameter.



Given an input parameter specified as optional \texttt{p}, this operation can be automated setting the CSP variable representing the presence of \texttt{p} to false (${pSet = false}$) and checking whether the problem has at least one solution. If it has no solutions, \texttt{p} is false optional. Note that the input \idl specification should be consistent, otherwise all parameters would be classified as false optional. This operation can be translated to a CSP as follows:


 \begin{align*}
 \mathrm{isFalse}&\mathrm{Optional(IDL,P,p)} \iff \\
 & \mathrm{isConsistentIDL(IDL,P)}~\wedge \\
 & \mathrm{solve(filter(\mathrm{map}(IDL,P),\{\{pSet,false\}\})) = \emptyset}
 \end{align*}


\noindent \textbf{\validSpecification}. This operation receives as input an \idl specification and the list of parameters included on it, and returns a Boolean indicating whether the specification is valid or not. An IDL specification is \emph{valid} if it is consistent (i.e., there exists at least one request satisfying all the dependencies of the specification) and it does not contain any dead or false optional parameters. This operation may be helpful to easily detect errors when editing service specifications. This operation can be translated to a CSP as follows:


\begin{align*}
\mathrm{isValid}&\mathrm{IDL(IDL,P)} \iff \mathrm{isConsistentIDL(IDL,P)}~\wedge \\
& \mathrm{\forall p_{i} \in P (\neg\xspace isDeadParameter(IDL,P,p_{i})} \; \\
& \mathrm{\wedge \neg\xspace isFalseOptional(IDL,P,p_{i})) \; }
\end{align*}



\noindent \textbf{\validRequest}. This operation takes as input an \idl specification, the full list of parameters from the API specification, and a service request (i.e., a list of parameters and their values) and returns a Boolean indicating whether the request is valid or not. A service request is valid if it satisfies all the dependencies of the \idl specification. This operation may be helpful for the early detection of invalid calls to the system. For example, an API gateway supporting this operation could detect invalid calls without the need to redirect the request to the service, providing faster responses and reducing the consumption of user quota. For example, the following is a valid request for the \idl specification depicted in Listing~\ref{list:valid-specification} : \texttt{\{p1=2,p2=5\}}.

\begin{lstlisting}[caption={Valid \idl specification.}, label={list:valid-specification}]
Or(p1, p2 AND p3);
OnlyOne(p2, p3);
\end{lstlisting}



Let \texttt{R} be an input request, i.e., a list of parameters and their respective values. This operation can be translated to a CSP by (1) setting the CSP variables related to each parameter to the value indicated in R, (2) setting the CSP variables related to the presence of the parameters in R to true (${R_{i}Set=true}$), (3) setting the CSP variables related to the parameters not included in R to false ($O_{i}Set=false$ where $O = P \setminus R$), and (4) checking whether the problem has at least one solution. If it has no solutions, it means that the request is not valid, namely:


\begin{align*}
\mathrm{is}&\mathrm{Valid}\mathrm{Request(IDL,P,R) \iff O = P \setminus R~\wedge} \\
&\mathrm{solve(filter(map(IDL,P), R~\cup }\\
& \mathrm{\{\{R_{1}Set,true\}, \{R_{2}Set,true\}, \ldots, \{R_{n}Set,true\}}\\
& \mathrm{\{O_{1}Set,false\}, \{O_{2}Set,false\}, \ldots, \{O_{k}Set,false\} \})) \neq \emptyset}
\end{align*}

\noindent \textbf{\validPartialRequest}. This operation is analogous to the previous one but the input request is \emph{partial} or incomplete, meaning that some other parameters should still be included to make it a full valid request. This operation returns a Boolean indicating whether the partial request is valid. A partial request is valid if it does not include any contradiction, i.e., it can be extended with new parameters to become a valid request. This operation may be helpful for the early detection of inconsistencies. For example, an interactive API documentation supporting this operation could warn the user about inconsistencies as soon as a dependency is violated, without having to wait until constructing the full request. 

Let \texttt{S} be a partial input request. This operation can be specified as a CSP as follows:


\begin{align*}
\mathrm{is}&\mathrm{Valid}\mathrm{PartialRequest(IDL,P,S)} \iff \\
&\mathrm{solve(filter(map(IDL,P), S~\cup }\\
& \mathrm{\{\{S_{1}Set,true\}, \{S_{2}Set,true\}, \ldots, \{S_{n}Set,true\}})) \neq \emptyset
\end{align*}




\noindent \textbf{\allRequests}. This operation receives as input the IDL specification of an API operation and the list of parameters from the API specification, and returns the list of all the possible requests to the service operation. As a precondition, all the parameters should have a discrete domain. Variants of this operation could be easily defined using standard combinatorial testing techniques, e.g., generate a list of requests that includes all the possible combinations of t parameters (t-wise testing~\cite{Nie2011}). This operation can be automated searching all the solutions of the corresponding CSP, namely:


\begin{align*}
\mathrm{allRequests(IDL,P)} = \mathrm{solveAll(\mathrm{map}(IDL,P))}
\end{align*}



\noindent \textbf{\numberRequests}. This operation also requires all the parameters to have a discrete domain. It takes as input the IDL specification of a service operation and the list of parameters from the API specification, and returns the total number of possible requests to the operation. This operation can be helpful to understand the size of the input space of a service. A large number of potential requests could indicate that the operation is too complex and that some refactoring is needed. This operation can be translated to CSP by simply getting the cardinality of the set of solutions, as shown below. It is worth mentioning, however, that CSP solvers often provide specific operations to calculate the number of solutions of a CSP more efficiently.

\begin{align*}
\mathrm{numberOfRequests(IDL,P)} = \mathrm{|solveAll(\mathrm{map}(IDL,P))|}
\end{align*}


\noindent \textbf{\randomRequest}. This operation receives as input the IDL specification of an API operation and the list of parameters from the API specification, and returns a random valid request for the operation. This operation, in combination with test data generators, may be useful for random testing of services, or as an initial step for the generation of more sophisticated test cases using search-based techniques, for example. This operation can be automated by translating the IDL specification to a CSP and finding a random solution, namely: 


\begin{align*}
\mathrm{randomRequest(IDL,P)} = \mathrm{solve(\mathrm{map}(IDL,P))}
\end{align*}

\section{Tooling Support}
\label{sec-tool}

As a part of our contribution, we provide a set of tools supporting the specification and analysis of inter-parameter dependencies in web APIs, including an editor of IDL specifications, an extension for the OAS language and an analysis library supporting the integration of our approach into any external project. Together, these components make our work readily applicable in practice and provide a reference implementation for future contributions on the topic.


\subsection{IDL Editor and Parser}
\label{sec-dsl}

\begin{figure}
  \centering
  \includegraphics[width=0.48\textwidth, cfbox=darkgray 0.1pt 0.1pt]{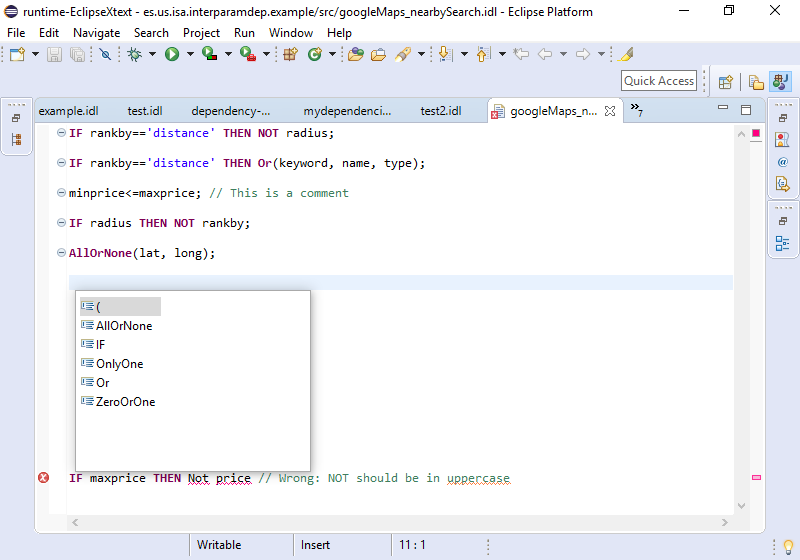}
  \caption{IDL Editor.} \label{fig:eclipse-editor}
\end{figure}

We implemented IDL using Xtext \cite{xtext}, a popular framework for the development of programming languages and DSLs. Xtext takes a grammar as input and generates a complete set of tools as output, including a linker, a compiler, a parser and a fully-fledged editor supporting features such as code completion, type checking, syntax coloring and validation. A simplified version of the IDL grammar is provided in Listing \ref{list:idl-grammar}, the full version is available as a part of the implementation of IDL \cite{idl} (and also on the supplemental material provided with this article \cite{landing-page}).
Figure~\ref{fig:eclipse-editor} depicts a screenshot of the editor, showing some of its capabilities: code completion, syntax coloring and error checking. The editor is based on Eclipse, but is compatible with any web browser or IDE supporting the Language Server Protocol \cite{lsp}.



\subsection{IDL4OAS: An OAS Extension}
\label{sec-idl4oas}


In order to foster the adoption of our approach, we propose an extension of OAS
for the specification of inter-parameter dependencies using IDL. We call this extension IDL4OAS.
An OAS document describes a REST API in terms of the elements it comprises, namely paths, operations, resources, request parameters and responses. OpenAPI provides a way to add extra information that may not be supported natively. This information is included in custom properties that start with ``x-'', called \emph{extensions}. IDL4OAS is an OAS extension that allows to specify a set of IDL dependencies for each API operation. An extra property called ``x-dependencies'' must be added at the operation level, including the set of dependencies among the input parameters of the operation. Listing \ref{list:idl4oas} shows an excerpt of an OAS document extended with IDL4OAS, corresponding to the \emph{Search} operation from the Google Maps Places API (see Listing \ref{list:googlemaps-idl}).


\lstset{
  language=YAML
}


\begin{lstlisting}[caption={OAS document of the search operation from the Google Maps Places API extended with IDL4OAS.}, label={list:idl4oas}]
paths:
  /search:
    get:
      parameters:
      - name: radius [...]
      - name: rankby [...]
      - name: keyword [...]
      - name: name [...]
      - name: type [...]
      - name: minprice [...]
      - name: maxprice [...]
      - [...]
      [...]
      x-dependencies:
      - ZeroOrOne(radius, rankby=='distance');
      - IF rankby=='distance' THEN keyword OR name OR type;
      - maxprice >= minprice;
\end{lstlisting}

As illustrated, the property ``x-dependencies'' has been added to the ``GET /search'' operation. This property is actually an array of elements, where each element represents a single dependency, therefore they must be preceded by hyphens, following the YAML syntax.

\subsection{IDLReasoner: An Analysis Library}
\label{sec-idlreasoner}
In this section, we present IDLReasoner, a CSP-based Java library that allows to programmatically analyse IDL documents. Specifically, IDLReasoner translates input \idl specifications to CSPs using MiniZinc \cite{minizinc}, a constraint solving language designed for modeling optimization problems in a high-level, solver-independent way. This allows IDLReasoner to be used with any CSP solver supporting MiniZinc as an input format.

\begin{figure}
  \centering
  \includegraphics[width=0.35\textwidth]{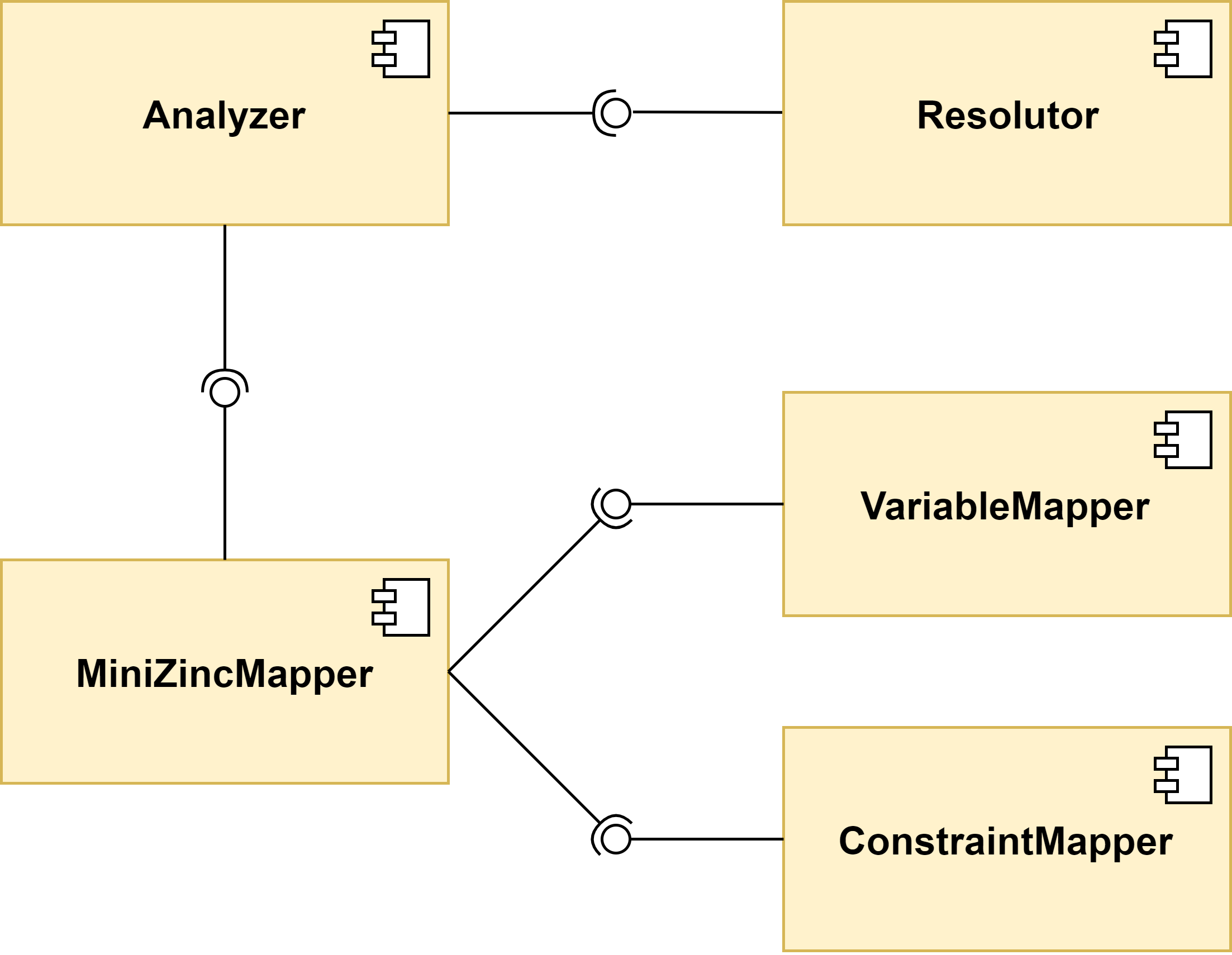}
  \caption{High-level architecture of IDLReasoner.} \label{fig:idlreasoner}
\end{figure}

Figure \ref{fig:idlreasoner} shows the high-level architecture of IDLReasoner, using a UML component diagram.
The library comprises three main components: the \emph{MiniZincMapper}, which translates variables from the API specification and dependencies from IDL to MiniZinc, and manipulates the resulting MiniZinc file accordingly for each analysis operation; the \emph{Resolutor}, which performs the calls to the selected CSP solver; and the \emph{Analyzer}, which leverages the MiniZincMapper and the Resolutor components to execute the \noperations analysis operations from the catalogue.

IDLReasoner works as follows: the Analyzer takes three elements as input, namely, an IDL document, an API specification (e.g., OAS) and the API operation where the dependencies are present (e.g., ``GET /search''). First, the MiniZincMapper transforms the API operation parameters, their domains and the IDL dependencies to a MiniZinc file, representing a CSP. Parameters and their domains are mapped by the \emph{VariableMapper} component, and dependencies are mapped to constraints by the \emph{ConstraintMapper} component. Then, when an analysis operation is invoked in the Analyzer component (e.g., valid specification), the MiniZincMapper manipulates the CSP file accordingly and the Resolutor calls the CSP solver on the manipulated file. IDLReasoner supports the \noperations analysis operations explained in Section \ref{sec-analysisOperations}. It is worth mentioning that IDLReasoner supports both IDL and OAS documents separately, as well as OAS documents including the specification of dependencies with IDL4OAS.





IDLReasoner is developed with extensibility in mind. It can be extended to multiple operating systems, web API specification languages and CSP solvers. At the time of writing this article, IDLReasoner supports Windows, OAS (and IDL4OAS), and a range of CSP solvers compatible with MiniZinc including Chuffed \cite{chuffed} and Gecode \cite{gecode}.

\section{Evaluation}
\label{sec-evaluation}


For the evaluation of our approach, we aim to answer two main research questions (RQs), namely:

\begin{itemize}
    \item \textbf{RQ1 - Expressiveness}. \textit{Is IDL expressive enough to model inter-parameter dependencies in real-world web APIs?} As previously mentioned, IDL lies on strong foundations since it is based on the dependencies observed in a large study of more than 2.5K operations of 40 industrial APIs. However, the expressiveness of the language to model real-world dependencies is still open to question and we aim to address this issue.
    
    \item \textbf{RQ2 - Validation}. \textit{Do the analysis operations implemented in IDLReasoner work as expected?} The implementation of the analysis operations is an error-prone process. Defects in the mapping to CSP and/or the involved tools could lead to unexpected functionality. By answering this question, we aim to assess that the analysis operations work as expected and to gain confidence in the correctness of the developed tools.
\end{itemize} 

\subsection{Expressiveness}
\label{sec-expressiveness}
To assess the expressiveness of \idl (RQ1), we evaluated the expresiveness of the language on the specification of inter-parameter dependencies in real-world APIs. Specifically, we selected 30 operations from 22 real-world APIs including Amazon Product Advertising, Box, GitHub, Groupon, PayPal, Shopify, Vimeo, and YouTube. A third part of the operations (10 out of 30) were taken from the APIs reviewed in our initial study of dependencies in web APIs, and the rest  were selected from APIs not previously studied. All selected APIs are ranked among the most popular APIs in ProgrammableWeb \cite{programmableweb}. In total, we specified 149 different dependencies using IDL covering the seven types of dependencies identified in our study.
As expected, we found no issues related to expressiveness and all dependencies were succinctly specified using the constructs of the language. This supports the suitability of \idl for the specification of dependencies among input parameters in practice. The IDL specifications are publicly available in the IDL repository \cite{idl}, as well as a brief description of the operations modelled and links to their corresponding API documentation.
\subsection{Validation}
\label{sec-validation}

For the validation of the developed tools (RQ2) we resorted to standard testing techniques. More specifically, we used equivalence partitioning, boundary-value analysis and combinatorial testing \cite{introductiontosoftwaretesting} for the construction of a test suite. In this scenario, the input domain is comprised of 
IDL4OAS specifications, individual parameters and API requests.
For the parameters, we used 3 partitions: valid, dead and false optional. For the requests, we used 2 partitions: valid and invalid. As for the IDL4OAS specifications,
we followed a combinatorial approach to cover all possible combinations of two different types of the following elements (so-called 2-wise testing or pairwise testing \cite{Nie2011}): number of parameters (5, 10), percentage of optional parameters (0, 50, 100), type of parameters (Booleans, integers, strings, enumerated integers, enumerated strings),
number of IDL dependencies (5, 10), type of IDL dependencies (\emph{Requires}, \emph{Or}, \emph{OnlyOne}, \emph{AllOrNone}, \emph{ZeroOrOne}, \emph{Arithmetic/Relational}, \emph{Complex}), and size of complex IDL dependencies (2, 5). The set of possible combinations was generated using the combinatorial testing tool PICT~\cite{pict}, developed by Microsoft.
We created additional specifications for testing the operations \emph{allRequests} and \emph{numberOfRequests}, since they require all parameters to have a finite domain.
In total, we designed and developed 178 test cases to test the functionality of both the editor and the analysis library. In the case of IDLReasoner, all test cases were automated using JUnit 5 \cite{junit5}.

As an example, Table \ref{tab:tc1} shows one of the cases for the operation \emph{isValidRequest}. The test case takes two inputs: a request and an IDL4OAS specification (Listing~\ref{list:idl4oas-test-case}). The operation is expected to return false since, according to the specification, requests including the parameter \texttt{p1} should include the parameters \texttt{p2} or \texttt{p3}, but not both at the same time.


\begin{table}[h]
    \centering
    \begin{tabular}{|p{5cm}|p{3cm}|}
        \hline
        \rowcolor[rgb]{0.8, 0.8, 0.8}
        \multicolumn{2}{|c|}{\textbf{Test case 1 - \emph{isValidRequest}}} \\
        \hline
        \thead{\textbf{Inputs}} & \thead{\textbf{Expected output}} \\
        \hline
        \textbf{Request}: \{p1=false, p2=`thing', p3=-10\}
         & \hspace{1.1cm} \multirow{2}{*}{\textbf{False}} \\
        \textbf{IDL4OAS}: Listing \ref{list:idl4oas-test-case} & \\
        \hline
    \end{tabular}
    \caption{Sample test case for the operation \emph{isValidRequest}.}
    \label{tab:tc1}
\end{table}

\lstset{
  language=YAML
}

\begin{lstlisting}[caption={IDL4OAS document used in several test cases.}, label={list:idl4oas-test-case}]
/oneDependency:
  get:
    parameters:
      - name: p1 [...]
      - name: p2 [...]
      - name: p3 [...]
      - name: p4 [...]
      - name: p5 [...]
    x-dependencies:
      - IF p1 THEN OnlyOne(p2, p3);
\end{lstlisting}

The test suite proved very useful to reveal failures in the developed tools, especially in IDLReasoner. Among other issues, fully documented in GitHub~\cite{idl, idlreasoner}, we detected and fixed faults related to the parsing of IDL specifications, their translation to MiniZinc files and the behaviour of the analysis operations for boundary cases such as operations without parameters. Once each issue was fixed, the suite was run again to make sure that no new defects had been introduced as a result of the changes, proving itself as a very effective tool for regression testing. Although performance is out of the scope of our work, it is worth mentioning that the execution of the whole suite took between 60 and 70 seconds in a standard PC running an Intel i5 processor with 16GB of RAM and a solid-state drive (SSD). As expected, most of the operations are run in a few milliseconds except the operations calculating all the possible requests of a specification, which are computationally expensive.

All the test cases, including their test inputs and expected outputs, are publicly available, as well as their implementation in JUnit~\cite{idlreasoner}. We trust that this data will be helpful not only for the sake of reproducibility, but also for future extensions of the tool suite or alternative implementations.

\section{Threats to Validity}
\label{sec-threats}
Next, we discuss the possible internal and external validity threats that may have influenced our work, and how these were mitigated.

\subsection{Internal Validity}
\label{sec-internal-validity}
Threats to the internal validity relate to those factors that might introduce bias and affect the results of our investigation. One of the main threats in this regard is the subjective and manual review process conducted for identifying inter-parameter dependencies in the online documentation of the subject APIs. Some dependencies might have been misclassified or simply overlooked. To mitigate this threat, the documentation of each API was carefully checked several times, recording all the relevant information for its later analysis, and also to enable replicability \cite{dataset}. The impact of possible mistakes was also minimised by the large number of APIs and operations reviewed (40 APIs and 2,557 operations), which makes us remain confident of the overall accuracy of the results.

Another possible threat is related to the existence of bugs in the implementation of the tools provided. To mitigate this threat, both the DSL and the analysis library have been thoroughly tested using standard testing techniques such as equivalence partitioning and combinatorial testing. During the development of the tool, several bugs were detected, either by manual inspection or by running the test suite (178 test cases). All bugs have been fixed and they are fully documented on GitHub \cite{idl, idlreasoner}. At the end of the development process, the test suite ran without any failures.

\subsection{External Validity}
\label{sec-external-validity}
This concerns the extent to which we can generalise from the results obtained in the experiments. Our study on the existence of inter-parameter dependencies in practice is based on a subset of 40 web APIs, and thus our results may not generalise to other APIs. To minimise this threat, we systematically selected a large set of real-world APIs from multiple application domains, including some of the most popular APIs in the world with millions of users worldwide.

As another threat, the DSL proposed in this paper could not be expressive enough to model all kinds of dependencies found in web APIs. However, several reasons make us confident in the expresiveness of the language. First, \idl is partially inspired by the grammar of PICT, a mature combinatorial testing tool developed by Microsoft. Second, \idl is based on the findings of a thorough study of over 600 dependencies found in more than 2.5K operations. Finally, and more importantly, we were able to model a total of 149 new dependencies from 22 real-world APIs, without identifying expresiveness issues.

Finally, our work lacks an empirical validation with software developers and practitioners that ensures the usefulness and usability of the developed tools. IDL might be considered hard to understand or to familiarise with. To minimise this threat, the language provides syntactic sugar to make dependencies self-explanatory (i.e., \emph{\dependencyOR}, \emph{\dependencyXOR}, \emph{\dependencyXNOR} and \emph{\dependencyNAND}). Also, we have proposed IDL4OAS, which allows to succinctly specify inter-parameter dependencies in OAS, the de-facto standard for API specification in industry.






\section{Related Work}
\label{sec-related}
Two related papers have addressed the issue of inter-parameter dependencies in web APIs. Wu et al.~\cite{wu13-www} presented an approach for the automated inference of dependency constraints among input parameters in web services. As a part of their work, they studied four popular web APIs and classified the dependencies found into six types, four of which are specific instances of the \emph{\dependencyREQUIRES} dependency presented in our work. Oostvogels et al.~\cite{Oostvogels2017323} proposed a DSL for the description of inter-parameter constraints in OAS. They first classified dependencies into three types: \emph{exclusive} (called \emph{\dependencyXOR} in our work), \emph{dependent} (\emph{\dependencyREQUIRES} in our work), and \emph{group constraints} (\emph{\dependencyXNOR} in our paper). Then, they looked for instances of those types of dependencies in the documentation of six popular APIs by searching for specific keywords such as ``either'' or ``one of''. Compared to theirs, our work is based on a much larger and systematic study: we have manually reviewed 40 APIs from different domains, whereas they have jointly studied 7 ``popular'' APIs. As a result, the conclusions drawn from our study differ from theirs. For example, we identified a richer set of dependencies (e.g., the DSL from Oostvogels et al. \cite{Oostvogels2017323} does not support all dependency types from our catalogue) and observed a different trend regarding the frequency of dependencies in real-world web APIs (e.g., Wu et al. \cite{wu13-www} found an average of 21.9\% of API operations to have dependencies, while in our study that percentage is 9.7\%). Furthermore, our work is the first to fully address both the specification and \emph{automated} analysis of inter-parameter dependencies.


In the context of RESTful web APIs, RAML \cite{raml} seems to be the only specification language that provides some support for the description of inter-parameter dependencies, albeit minimal. Mutually exclusive parameters (i.e., \emph{\dependencyXOR} dependencies from our catalogue) can be specified with the so-called \emph{union} type, where a piece of data can be described by any of several types. For example, to describe an operation with one required parameter \texttt{p1} and two mutually exclusive parameters \texttt{p2} and \texttt{p3}, it could be done as follows: ``\texttt{queryString: type: [p1, p2 | p3]}''. However, RAML does not offer support for the remaining six dependency types presented in this article, which represent 83\% of the dependencies found in our study of real-world web APIs \cite{alberto19icsoc}.

Other than RESTful services, several authors \cite{Xu200959, Gao201465, Cacciagrano2006138} have partially addressed the specification, inference or analysis of inter-parameter dependencies in other types of web services such as WSDL \cite{wsdl} and OWL-S \cite{owls}, technologies  increasingly in disuse nowadays. Compared to them, our approach is specification-independent and is based on the first large-scale study of inter-parameter dependencies in web APIs \cite{alberto19icsoc}.

Regarding the specification of dependencies, combinatorial test case generation tools offer similar capabilities to specify constraints among input parameters, e.g., TestCover \cite{testcover}, Advanced Combinatorial Testing System (ACTS) \cite{ACTS2013} and Pairwise Independent Combinatorial Testing (PICT) \cite{pict}. Unfortunately, these tools were not designed with reusability in mind and their use out of the context of testing is hardly possible. The syntax of IDL is partially based on that of PICT, a fully-fledged tool developed by Microsoft. However, we extended the constraints grammar of PICT to support the seven types of dependencies from our catalogue \cite{alberto19icsoc}, and to provide syntactic sugar that makes dependencies succinct and self-explanatory. 

Regarding the automated analysis of \idl specifications, our proposal is inspired by previous work by the authors in the context of feature models, where more than 30 different analysis operations have been proposed~\cite{benavides10}. Also, we were pioneers on the automated analysis of service level agreements in different web service technologies such as WS-Agreement \cite{Muller13TSC, Muller18TSC, Muller14}, Linked USDL \cite{Garcia17} and recently in OAS \cite{GamezDiaz19}. Both lines of research have served as the basis for the contributions presented in this paper where, although with similar ideas, we had to face unique challenges due to the richer catalogue of dependency patterns found in web APIs.





\section{Conclusions}
\label{sec-conclusions}

This article addressed the problem of specifying the dependencies among input parameters in web APIs. We presented a domain specific language, \idl, specifically designed to express the seven types of dependencies observed in a thorough study of more than 2.5K operations from 40 industrial APIs. Besides this, we proposed a catalogue of \noperations analysis operations to extract helpful information from \idl specifications such as detecting inconsistencies or checking the validity of API requests. For the automation of the analysis operations we proposed translating \idl specifications to CSPs and leveraging the capabilities of state-of-the-art CSP solvers. The approach is supported by an (Eclipse) editor, a parser, an OAS extension, and an analysis library validated through a thorough test suite of \ntestcases test cases. Together, these contributions not only provide a complete solution to the automated management of inter-parameter dependencies in web APIs, but they also open a new range of applications and research opportunities in areas such as code generation, monitoring and testing.

There are a number of possible lines of future work. First, new analysis operations could be defined to further leverage \idl specifications. For example, the explanation feature integrated in most CSP solvers could be used to explain the causes of inconsistencies, e.g., why a given parameter is dead. Second, the integration of \idl in other API specification languages like RAML would facilitate its adoption and would further test the generalisability of our approach. Last, but not least, our contributions pave the way for new promising applications by leveraging both the specification of dependencies and their automated analyses.

\section*{Verifiability}
For the sake of verifiability, we provide an online appendix pointing to the resources used in this article~\cite{landing-page}: (1) links to the GitHub repositories of IDL and IDLReasoner, (2) dataset of our study on the presence of inter-parameter dependencies on real-world web APIs, (3) links to the documentation of the web APIs involved in the study and in the expressiveness assessment (RQ1), (4) test suite used for the validation of our approach (RQ2), and (5) demo video showing the capabilities of the IDL editor and the analysis library.

\ifCLASSOPTIONcompsoc
  \section*{Acknowledgements}
\else
  \section*{Acknowledgement}
\fi
This work has been partially supported by the European Commission (FEDER) and Spanish Government under projects APOLO (US-1264651) and HORATIO (RTI2018-101204-B-C21), and the FPU scholarship program, granted by the Spanish Ministry of Education and Vocational Training (FPU17/04077). We would also like to thank Roberto Hermoso for his technical support during the development of IDLReasoner.

\ifCLASSOPTIONcaptionsoff
  \newpage
\fi

\bibliographystyle{IEEEtranS}
\bibliography{segura}

\begin{thebibliography}{10}
\providecommand{\url}[1]{#1}
\csname url@samestyle\endcsname
\providecommand{\newblock}{\relax}
\providecommand{\bibinfo}[2]{#2}
\providecommand{\BIBentrySTDinterwordspacing}{\spaceskip=0pt\relax}
\providecommand{\BIBentryALTinterwordstretchfactor}{4}
\providecommand{\BIBentryALTinterwordspacing}{\spaceskip=\fontdimen2\font plus
\BIBentryALTinterwordstretchfactor\fontdimen3\font minus
  \fontdimen4\font\relax}
\providecommand{\BIBforeignlanguage}[2]{{%
\expandafter\ifx\csname l@#1\endcsname\relax
\typeout{** WARNING: IEEEtranS.bst: No hyphenation pattern has been}%
\typeout{** loaded for the language `#1'. Using the pattern for}%
\typeout{** the default language instead.}%
\else
\language=\csname l@#1\endcsname
\fi
#2}}
\providecommand{\BIBdecl}{\relax}
\BIBdecl

\bibitem{introductiontosoftwaretesting}
P.~Ammann and J.~Offutt, \emph{Introduction to Software Testing}.\hskip 1em
  plus 0.5em minus 0.4em\relax Cambridge University Press, 2016.

\bibitem{benavides10}
{Benavides, D., Segura, S., Ruiz-Cortés, A.}, ``{{Automated Analysis of
  Feature Models 20 Years Later: A Literature Review}},'' \emph{{Information
  Systems}}, vol.~{35}, no.~{6}, pp. {615 -- 636}, {2010}.

\bibitem{bing-api}
\BIBentryALTinterwordspacing
``{Bing Web Search API},'' accessed January 2020. [Online]. Available:
  \url{https://docs.microsoft.com/en-us/rest/api/cognitiveservices/bing-web-api-v7-reference}
\BIBentrySTDinterwordspacing

\bibitem{bing-maps-api}
\BIBentryALTinterwordspacing
``{Bing Maps API},'' accessed January 2020. [Online]. Available:
  \url{https://msdn.microsoft.com/en-us/library/ff701702.aspx}
\BIBentrySTDinterwordspacing

\bibitem{Cacciagrano2006138}
D.~Cacciagrano, F.~Corradini, R.~Culmone, and L.~Vito, ``{Dynamic
  Constraint-based Invocation of Web Services},'' in \emph{3rd Intern. Workshop
  on Web Services and Formal Methods}, 2006, pp. 138--147.

\bibitem{chuffed}
\BIBentryALTinterwordspacing
``{The Chuffed CP solver},'' accessed January 2020. [Online]. Available:
  \url{https://github.com/chuffed/chuffed}
\BIBentrySTDinterwordspacing

\bibitem{dataset}
\BIBentryALTinterwordspacing
``{Inter-Parameter Dependencies in RESTful APIs [Dataset]},'' 2019. [Online].
  Available: \url{https://bit.ly/2wvv1m1}
\BIBentrySTDinterwordspacing

\bibitem{fielding00-phd}
R.~T. Fielding, ``{Architectural Styles and the Design of Network-based
  Software Architectures},'' Ph.D. dissertation, 2000.

\bibitem{flickr-api}
\BIBentryALTinterwordspacing
``{Flickr API},'' accessed January 2020. [Online]. Available:
  \url{https://www.flickr.com/services/api/}
\BIBentrySTDinterwordspacing

\bibitem{forte-api}
\BIBentryALTinterwordspacing
``{Forte API},'' accessed January 2020. [Online]. Available:
  \url{https://restdocs.forte.net/}
\BIBentrySTDinterwordspacing

\bibitem{foursquare-api}
\BIBentryALTinterwordspacing
``{Foursquare API},'' accessed January 2020. [Online]. Available:
  \url{https://developer.foursquare.com/places-api}
\BIBentrySTDinterwordspacing

\bibitem{GamezDiaz19}
A.~Gamez-Diaz, P.~Fernandez, and A.~Ruiz-Cort\'{e}s, ``{Governify for APIs:
  SLA-Driven Ecosystem for API Governance},'' in \emph{ACM Joint Meeting on
  European Software Engineering Conference and Symposium on the Foundations of
  Software Engineering}, 2019, p. 1120–1123.

\bibitem{Gao201465}
C.~Gao, J.~Wei, H.~Zhong, and T.~Huang, ``{Inferring Data Contract for
  Web-based API},'' in \emph{IEEE Intern. Conference on Web Services}, 2014,
  pp. 65--72.

\bibitem{Garcia17}
J.~Garc{\'i}a, P.~Fernandez, C.~Pedrinaci, M.~Resinas, J.~Cardoso, and
  A.~Ruiz-Cort{\'e}s, ``{Modeling Service Level Agreements with Linked USDL
  Agreement},'' \emph{IEEE Transactions on Services Computing}, vol.~10, no.~1,
  pp. 52--65, 2017.

\bibitem{gecode}
\BIBentryALTinterwordspacing
``{GECODE - An open, free, efficient constraint solving toolkit},'' accessed
  January 2020. [Online]. Available: \url{https://www.gecode.org/}
\BIBentrySTDinterwordspacing

\bibitem{geonames-api}
\BIBentryALTinterwordspacing
``{GeoNames API},'' accessed January 2020. [Online]. Available:
  \url{http://www.geonames.org}
\BIBentrySTDinterwordspacing

\bibitem{github-api}
\BIBentryALTinterwordspacing
``{GitHub API},'' accessed January 2020. [Online]. Available:
  \url{https://developer.github.com/v3/}
\BIBentrySTDinterwordspacing

\bibitem{google-maps-api}
\BIBentryALTinterwordspacing
``{Google Maps API},'' accessed January 2020. [Online]. Available:
  \url{https://developers.google.com/places/web-service/intro}
\BIBentrySTDinterwordspacing

\bibitem{idl}
\BIBentryALTinterwordspacing
``{Inter-parameter Dependency Language (IDL)},'' accessed January 2020.
  [Online]. Available: \url{https://github.com/isa-group/IDL}
\BIBentrySTDinterwordspacing

\bibitem{idlreasoner}
\BIBentryALTinterwordspacing
``{IDLReasoner: An Analysis Library for IDL Specifications},'' accessed January
  2020. [Online]. Available: \url{https://github.com/isa-group/IDLReasoner}
\BIBentrySTDinterwordspacing

\bibitem{jacobson11-book}
D.~Jacobson, G.~Brail, and D.~Woods, \emph{{APIs: A Strategy Guide}}.\hskip 1em
  plus 0.5em minus 0.4em\relax O'Reilly Media, Inc., 2011.

\bibitem{jacobson14-oscon}
\BIBentryALTinterwordspacing
D.~Jacobson and S.~Narayanan, ``{Netflix API: Top 10 Lessons Learned},'' in
  \emph{Open Source Convention (OSCON)}, Porland, Oregon, July 2014. [Online].
  Available:
  \url{http://www.slideshare.net/danieljacobson/top-10-lessons-learned-from-the-netflix-api-oscon-2014}
\BIBentrySTDinterwordspacing

\bibitem{junit5}
\BIBentryALTinterwordspacing
``{JUnit 5},'' accessed January 2020. [Online]. Available:
  \url{http://junit.org/junit5/}
\BIBentrySTDinterwordspacing

\bibitem{lastfm-api}
\BIBentryALTinterwordspacing
``{Last.fm API},'' accessed January 2020. [Online]. Available:
  \url{https://www.last.fm/api}
\BIBentrySTDinterwordspacing

\bibitem{lsp}
\BIBentryALTinterwordspacing
``{Language Server Protocol},'' accessed January 2020. [Online]. Available:
  \url{https://microsoft.github.io/language-server-protocol}
\BIBentrySTDinterwordspacing

\bibitem{alberto19icsoc}
A.~Martin-Lopez, S.~Segura, and A.~Ruiz-Cort{\'e}s, ``{A Catalogue of
  Inter-Parameter Dependencies in RESTful Web APIs},'' in \emph{Intern.
  Conference on Service-Oriented Computing}, 2019, pp. 399--414.

\bibitem{minizinc}
\BIBentryALTinterwordspacing
``{MiniZinc: Constraint Modeling Language},'' accessed November 2019. [Online].
  Available: \url{https://www.minizinc.org/}
\BIBentrySTDinterwordspacing

\bibitem{Muller14}
C.~Müller, M.~Oriol, X.~Franch, J.~Marco, M.~Resinas, A.~Ruiz-Cort{\'e}s, and
  M.~Rodr{\'i}guez, ``{Comprehensive Explanation of SLA Violations at
  Runtime},'' \emph{IEEE Transactions on Services Computing}, vol.~7, no.~2,
  pp. 168--183, 2014.

\bibitem{Muller18TSC}
C.~M{\"u}ller, A.~M. Gutierrez~Fernandez, P.~Fernandez, O.~Martin-Diaz,
  M.~Resinas, and A.~Ruiz-Cortes, ``{{A}utomated {V}alidation of {C}ompensable
  {SLA}s},'' \emph{IEEE Transactions on Services Computing}, 2018, article in
  press.

\bibitem{Muller13TSC}
C.~M{\"u}ller, M.~Resinas, and A.~Ruiz-Cortés, ``{{A}utomated {A}nalysis of
  {C}onflicts in {WS--A}greement},'' \emph{IEEE Transactions on Services
  Computing}, vol.~7, no.~4, pp. 530--544, 2014.

\bibitem{nationbuilder-api}
\BIBentryALTinterwordspacing
``{NationBuilder API},'' accessed January 2020. [Online]. Available:
  \url{https://nationbuilder.com/api_documentation}
\BIBentrySTDinterwordspacing

\bibitem{Nie2011}
C.~Nie and H.~Leung, ``{A Survey of Combinatorial Testing},'' \emph{ACM
  Computing Surveys}, vol.~43, no.~2, 2011.

\bibitem{Oostvogels2017323}
{Oostvogels, N., De Koster, J., De Meuter, W.}, ``{{Inter-parameter Constraints
  in Contemporary Web APIs}},'' in \emph{{17th Intern. Conference on Web
  Engineering}}, {2017}, pp. {323--335}.

\bibitem{oai}
\BIBentryALTinterwordspacing
``{OpenAPI Specification},'' accessed March 2019. [Online]. Available:
  \url{https://github.com/OAI/OpenAPI-Specification}
\BIBentrySTDinterwordspacing

\bibitem{owls}
\BIBentryALTinterwordspacing
``{Semantic Markup for Web Services (OWL-S)},'' accessed November 2019.
  [Online]. Available: \url{https://www.w3.org/Submission/OWL-S/}
\BIBentrySTDinterwordspacing

\bibitem{pict}
\BIBentryALTinterwordspacing
``{Microsoft PICT - Pairwise Independent Combinatorial Testing},'' accessed
  October 2019. [Online]. Available: \url{https://github.com/microsoft/pict}
\BIBentrySTDinterwordspacing

\bibitem{programmableweb}
\BIBentryALTinterwordspacing
``{ProgrammableWeb API Directory},'' accessed March 2019. [Online]. Available:
  \url{http://www.programmableweb.com/}
\BIBentrySTDinterwordspacing

\bibitem{quickbooks-api}
\BIBentryALTinterwordspacing
``{QuickBooks Payments API},'' accessed January 2020. [Online]. Available:
  \url{https://developer.intuit.com/app/developer/qbpayments/docs/get-started/}
\BIBentrySTDinterwordspacing

\bibitem{raml}
\BIBentryALTinterwordspacing
``{RESTful API Modeling Language (RAML)},'' accessed March 2019. [Online].
  Available: \url{http://raml.org/}
\BIBentrySTDinterwordspacing

\bibitem{rapidapi}
\BIBentryALTinterwordspacing
``{RapidAPI API Directory},'' accessed March 2019. [Online]. Available:
  \url{https://rapidapi.com}
\BIBentrySTDinterwordspacing

\bibitem{richardson13-book}
L.~Richardson, M.~Amundsen, and S.~Ruby, \emph{{RESTful Web APIs}}.\hskip 1em
  plus 0.5em minus 0.4em\relax O'Reilly Media, Inc., 2013.

\bibitem{stripe-api}
\BIBentryALTinterwordspacing
``{Stripe API},'' accessed January 2020. [Online]. Available:
  \url{https://stripe.com/docs/api}
\BIBentrySTDinterwordspacing

\bibitem{landing-page}
\BIBentryALTinterwordspacing
``{Supplementary material of the paper}.'' [Online]. Available:
  \url{https://isa-group.github.io/2020-02-inter-parameter-dependencies}
\BIBentrySTDinterwordspacing

\bibitem{swagger}
\BIBentryALTinterwordspacing
``{Swagger},'' accessed March 2019. [Online]. Available:
  \url{http://swagger.io/}
\BIBentrySTDinterwordspacing

\bibitem{Hofstede98}
A.~H.~M. Ter~Hofstede and H.~A. Proper, ``{{H}ow to {F}ormalize {I}t?
  {F}ormalization {P}rinciples for {I}nformation {S}ystem {D}evelopment
  {M}ethods},'' \emph{Information and Software Technology}, vol.~40, pp.
  519--540, 1998.

\bibitem{testcover}
\BIBentryALTinterwordspacing
``{Testcover.com},'' accessed January 2020. [Online]. Available:
  \url{https://www.testcover.com/}
\BIBentrySTDinterwordspacing

\bibitem{twilio-api}
\BIBentryALTinterwordspacing
``{Twilio API},'' accessed January 2020. [Online]. Available:
  \url{https://www.twilio.com/docs/usage/api/}
\BIBentrySTDinterwordspacing

\bibitem{wsdl}
\BIBentryALTinterwordspacing
``{Web Services Description Language (WSDL) Version 2.0},'' accessed November
  2019. [Online]. Available: \url{https://www.w3.org/TR/wsdl20/}
\BIBentrySTDinterwordspacing

\bibitem{wu13-www}
Q.~Wu, L.~Wu, G.~Liang, Q.~Wang, T.~Xie, and H.~Mei, ``{Inferring Dependency
  Constraints on Parameters for Web Services},'' in \emph{Proceedings of the
  22nd Intern. Conference on World Wide Web}, 2013, pp. 1421--1432.

\bibitem{xtext}
\BIBentryALTinterwordspacing
``{Xtext},'' accessed October 2019. [Online]. Available:
  \url{https://www.eclipse.org/Xtext/index.html}
\BIBentrySTDinterwordspacing

\bibitem{Xu200959}
L.~Xu, Q.~Yuan, J.~Wu, and C.~Liu, ``{Ontology-based Web Service Robustness
  Test Generation},'' in \emph{IEEE Intern. Symp. on Web Systems Evolution},
  2009, pp. 59--68.

\bibitem{yelp-api}
\BIBentryALTinterwordspacing
``{Yelp API},'' accessed January 2020. [Online]. Available:
  \url{https://www.yelp.com/developers/documentation/v3}
\BIBentrySTDinterwordspacing

\bibitem{twitter-api}
\BIBentryALTinterwordspacing
``{Twitter API},'' accessed January 2020. [Online]. Available:
  \url{https://developer.twitter.com/en/docs}
\BIBentrySTDinterwordspacing

\bibitem{youtube-api}
\BIBentryALTinterwordspacing
``{YouTube Data API v3},'' accessed January 2019. [Online]. Available:
  \url{https://developers.google.com/youtube/v3/}
\BIBentrySTDinterwordspacing

\bibitem{ACTS2013}
L.~{Yu}, Y.~{Lei}, R.~N. {Kacker}, and D.~R. {Kuhn}, ``{ACTS: A Combinatorial
  Test Generation Tool},'' in \emph{International Conference on Software
  Testing, Verification and Validation}, 2013, pp. 370--375.

\end{thebibliography}

\vspace{-9cm}

\begin{IEEEbiography}[{\includegraphics[width=1in,height=1.25in,clip,keepaspectratio]{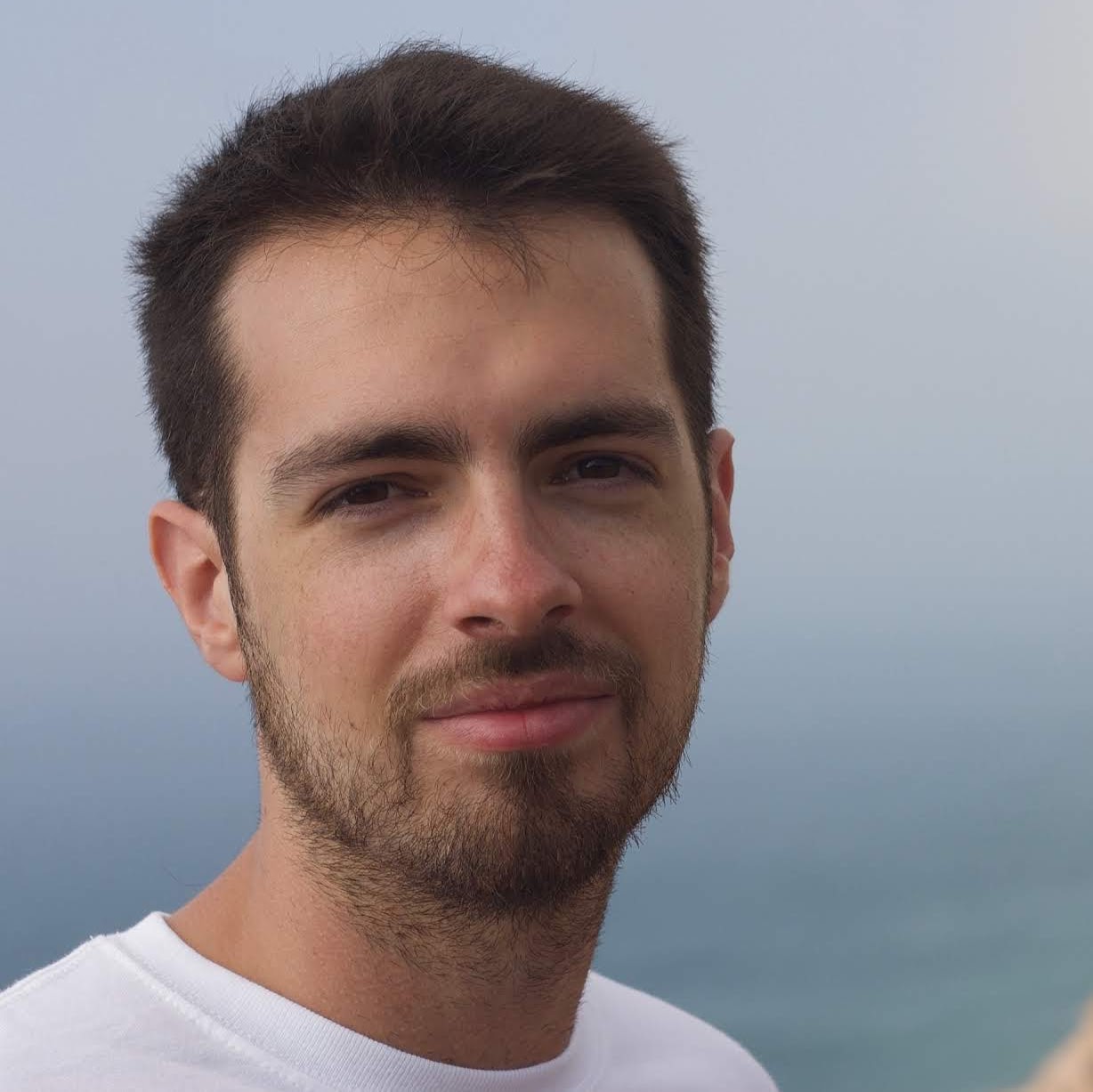}}]{Alberto Martin-Lopez} is a PhD candidate at the Applied Software Engineering research group (ISA, www.isa.us.es), University of Seville, Spain. He received his MsC from this university. His current research interests focus on automated software testing and service-oriented architectures.
\end{IEEEbiography}

\vskip -2\baselineskip plus -1fil

\begin{IEEEbiography}[{\includegraphics[width=1in,height=1.25in,clip,keepaspectratio]{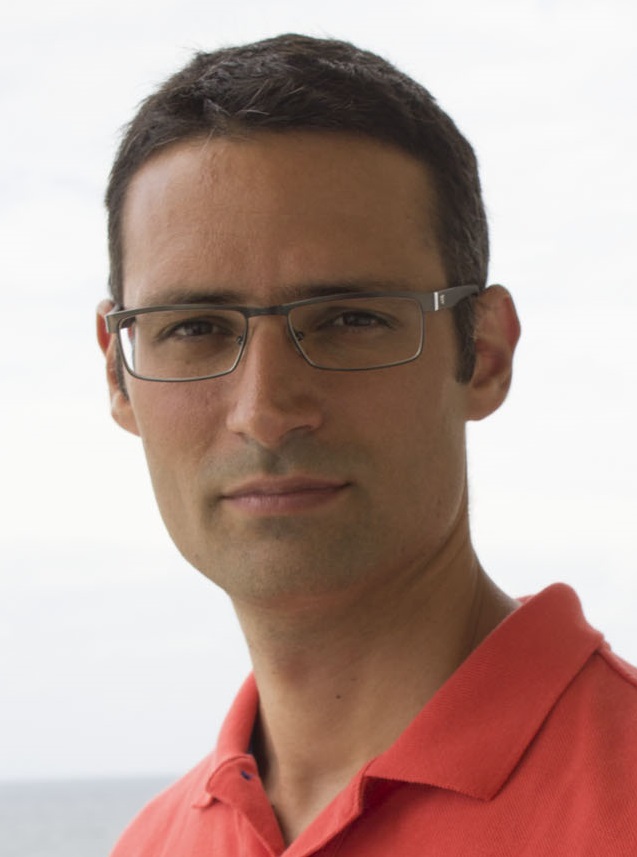}}]{Sergio Segura} is an Associate Professor of software engineering at the University of Seville, Spain. He is a member of the Applied Software Engineering research group, where he leads the research lines on software testing and search-based software engineering. His current research interests include test automation and AI-driven software engineering. Contact him at sergiosegura@us.es.
\end{IEEEbiography}

\vskip -2\baselineskip plus -1fil

\begin{IEEEbiography}[{\includegraphics[width=1in,height=1.25in,clip,keepaspectratio]{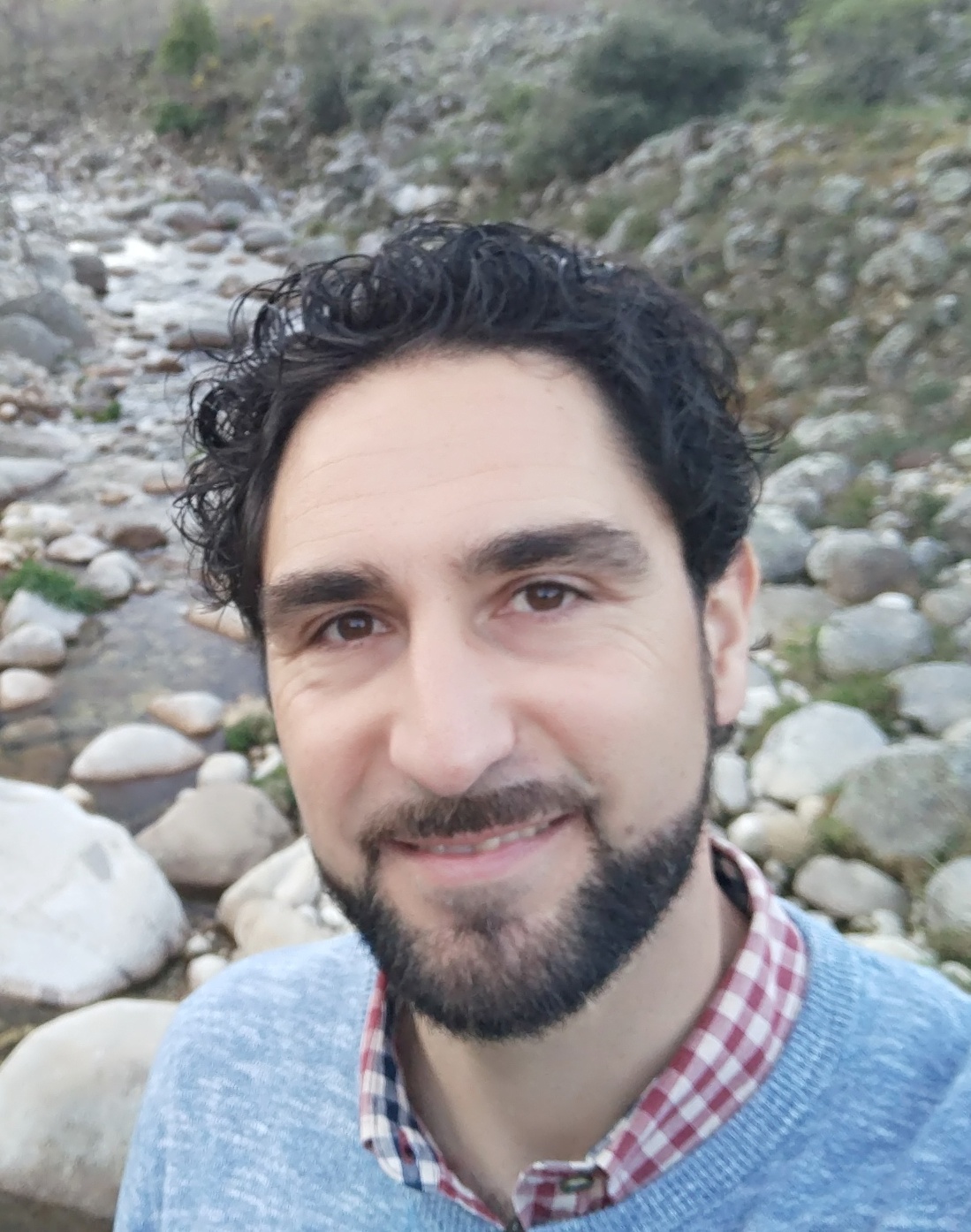}}]{Carlos M{\"u}ller}
is a Lecturer and member of the Applied Software Engineering Group (ISA, www.isa.us.es) at University of Sevilla, Spain. He obtained his PhD in Computer Science from this university. His current research line includes the automated analysis of service level agreements (SLA) and the application of such analysis at SLA design and monitoring.
\end{IEEEbiography}

\vskip -2\baselineskip plus -1fil

\begin{IEEEbiography}[{\includegraphics[width=1in,height=1.25in,clip,keepaspectratio]{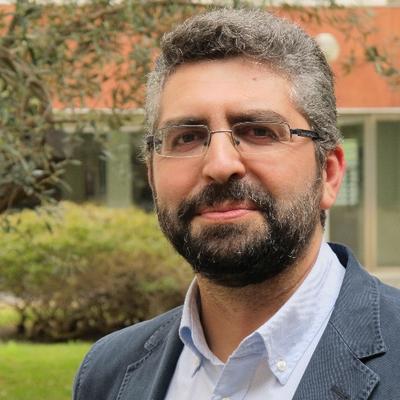}}]{Antonio Ruiz-Cort{\'e}s} is a Full Professor of software and service engineering and he heads the Applied Software Engineering Group at the University of Sevilla. His current research focuses on service-oriented computing, business process management, testing and software product lines, being the recipient of the Most Influential Paper of SPLC 2017 award. He is an associate editor of Springer Computing. Contact him at aruiz@us.es.
\end{IEEEbiography}









\end{document}